\documentclass[letterpaper,draftcls,onecolumn,journal,12pt,dvipdfmx]{IEEEtran} 
\usepackage{graphicx,amsmath,amssymb,cite,bm,balance,cases,multirow,color}
\usepackage{algorithm,algorithmic}
\renewcommand{\algorithmicrequire}{\textbf{Input:}}
\renewcommand{\algorithmicensure}{\textbf{Output:}}

\newtheorem{theorem}{Theorem}
\newtheorem{lemma}{Lemma}
\newtheorem{remark}{Remark}

\begin{document}

\title{Capacity-Approaching Polar Codes with 
Long Codewords and Successive Cancellation Decoding
Based on Improved Gaussian Approximation}

\author{Hideki Ochiai, \IEEEmembership{Senior Member, IEEE}, 
Patrick Mitran, \IEEEmembership{Senior Member, IEEE}, and \\
H. Vincent Poor, \IEEEmembership{Fellow, IEEE}%
\thanks{H. Ochiai is with the Department of Electrical and Computer Engineering, 
Yokohama National University, Yokohama, Japan. email: hideki@ynu.ac.jp}
\thanks{P. Mitran is with the Department of Electrical and Computer Engineering, 
University of Waterloo, ON, Canada. email:mitran@uwaterloo.ca}
\thanks{H. V. Poor is with the Department of Electrical Engineering, 
Princeton University, Princeton, NJ. email: poor@princeton.edu}\thanks{%
This work was supported in part by 
the Japan Society for the Promotion of Science (JSPS) through the Grants-in-Aid for Scientific Research (KAKENHI) under Grant JP16KK0145, in part by the Natural Sciences and Engineering Research Council of Canada (NSERC), and in part by the U.S. National Science Foundation under Grant CCF-1908308.}
}
\date{}

\maketitle

\vspace{-1.8cm}

\begin{abstract}
This paper focuses on an improved Gaussian approximation~(GA) based construction of polar codes with successive cancellation~(SC) decoding over an additive white Gaussian noise~(AWGN) channel.
Ar{\i}kan 
has proven that polar codes with low-complexity SC decoder can approach the channel capacity of an arbitrary symmetric binary-input discrete memoryless channel, provided that the code length is chosen large enough. Nevertheless, how to construct such codes over an AWGN channel with low computational effort has been an open problem. 
Compared to density evolution,
the GA is known as a low complexity yet powerful 
technique that traces the evolution of the mean log likelihood ratio~(LLR) value 
by iterating a nonlinear function. Therefore, its high-precision numerical evaluation
is critical as the code length increases. In this work, by analyzing the asymptotic behavior of this nonlinear function, we propose an improved GA approach that makes an accurate trace of mean LLR evolution feasible.
With this improved GA, through numerical analysis and simulations with code lengths up to $N=2^{18}$, we explicitly demonstrate that various code-rate polar codes with long codeword and capacity approaching behavior can be easily designed.

\end{abstract}

\vspace{-.6cm}


\IEEEpeerreviewmaketitle

\section{Introduction}

Polar codes, introduced by Ar{\i}kan~\cite{arikan09},
have the salient property that they can 
achieve channel capacity by low-complexity successive cancellation~(SC) decoding.
In order to approach capacity over a practical additive white Gaussian noise~(AWGN) channel
by SC decoding, not only must the polar code length be large, but 
the code
structure must also be properly designed.
A key difficulty of polar code design over binary-input AWGN channels
lies in the fact that the optimal polar code that yields
the lowest block error rate~(BLER) 
by SC decoding 
may have different code structure depending
on the specific channel signal-to-noise power ratio~(SNR).
Therefore, practical code design should be able to predict its BLER performance
for given operating SNR and code rate. 

In the framework of polar codes, 
code construction is equivalent to the selection of 
information bit locations, collectively referred to as the {\em information set}~\cite{arikan09}.
These are chosen from among all the input bits to the rate-1 polar encoder.
Several approaches for polar code construction, i.e.,
algorithms for information set selection,
have been developed in the literature for an AWGN channel. 
Mori and Tanaka~\cite{mori2009} proposed
the use of density evolution~(DE), originally developed for the design of
irregular 
low-density parity-check~(LDPC) codes by
Richardson {\em et al.} in~\cite{richardson2001}.
Since memory and computation 
requirements
for 
direct tracing of density 
grow exponentially 
with code length,
Tal and Vardy developed a tractable approach that
manages the memory requirements
by efficiently merging the density~\cite{tal2013}
through quantization at the cost of a loss in channel capacity.  
The resulting algorithms (with channel upgrading and degrading
due to quantization) were shown to provide 
rigorous upper and lower bounds 
to the achievable performance, 
and their tightness depends on the number of quantization levels~$\mu$.
Upon merging the density, 
the sorting of likelihood ratios with memory size of order $O(\mu^2)$ is required,
and 
the overall complexity of this is of order 
$O(N \mu^2 \log \mu)$ for polar codes with codeword length $N$~\cite{tal2013}.
In order to 
tighten
the bounds, one should increase $\mu$
without sacrificing numerical accuracy when calculating the loss 
in channel capacity.
This makes the algorithm 
computationally challenging, especially
for the design of polar codes with large block lengths under various operating SNR.
Other constructions include 
reduced complexity
DE based on the sub-optimal min-sum algorithm~\cite{Kern2014},
which still imposes a memory requirement that grows exponentially
as the codeword length increases.

Meanwhile, for design and analysis of LDPC codes,
Chung {\em et al.} proposed 
the use of a Gaussian approximation~(GA)~\cite{chung2001}.
Instead of tracing the exact density of the log likelihood ratios~(LLR), GA
only traces the mean value.
This becomes a sufficient statistic to characterize the LLR
provided that 
the LLR is assumed to be modeled, for some $\gamma > 0$,
as a Gaussian distribution with mean $\gamma$
and variance $2\gamma$. Since the GA method can be implemented with only 
a few computations (that involve nonlinear functions), 
the overall complexity 
is 
proportional to the codeword length $N$.
Trifonov~\cite{trifonov2012} demonstrated that the GA can be also used
for constructing polar codes with SC decoding.
Subsequently, the effectiveness of the GA for polar code construction
has been confirmed in~\cite{wu2014,li_yuan2013}.
Furthermore, in~\cite{li_yuan2013,Vangala2015},
the use of the Bhattacharrya parameter~\cite{arikan09}
for information set selection,
which is strictly valid only for the case of binary erasure channels,
has been shown by simulation to be effective provided that
the operating SNR parameter is appropriately chosen by an empirical approach.

In order for polar codes with SC decoding to outperform LDPC codes,
it is known that the code length should be much larger than that of LDPC codes~\cite{Korada2010}.
In other words, polar codes over an AWGN channel 
with moderate code length
are not competitive with other capacity approaching codes.
Since then, the use of other decoding approaches 
such as belief propagation~(BP) decoding
has also been studied~\cite{Arikan2008bp,Hussami2009,Eslami2013}.
On the other hand,
Tal and Vardy showed that successive cancellation {\em list} (SCL) decoding 
improves the performance, and furthermore, the application of error detection
codes such as cyclic redundancy-check~(CRC) codes in combination with
SCL decoding is shown to outperform other competitive codes~\cite{tal2015}.
Subsequently, most studies on polar codes target moderate block-length codes
assuming more sophisticated decoding. As a consequence, studies 
for polar code construction have also pursued decoding 
specific designs such as those proposed in~\cite{Qin2017} and~\cite{Elkelesh2019}.

Here, we recall that a major advantage of polar codes as proposed by 
Ar{\i}kan~\cite{arikan09}
is its simplicity of encoding and SC decoding with its capacity achieving performance.
The only issue is that the code length should be much larger
compared to LDPC codes~\cite{Korada2010} for the same or better target performance.
Therefore, 
this paper focuses on 
constructing polar codes for SC decoding with very large 
code lengths (i.e., larger than $10^5$ bits) and evaluating their
performance over a practical AWGN channel 
based on GA as a low-complexity design approach
compared to the original DE.
Furthermore, the assumption of 
GA enables us to derive an estimate of 
the BLER 
without resorting to time-consuming simulations.
In this work, 
we first elucidate the limitation of the conventional
GA in its mean LLR estimation process,
especially as the code length increases. 
It turns out that the straightforward implementation 
of conventional GA fails to construct good polar codes with capacity approaching behavior
at large block length
(as will be illustrated in Fig.~\ref{fig:BLER_CONV_GA}
in Section~\ref{subsec:comparison}).
This stems from the fact that the dynamic range of the LLR values
increases exponentially with code length, and thus the conventional approach
cannot accurately trace the evolution of mean LLR values.
We then propose a solution to this numerical issue by carefully
studying the behavior of specific functions involved
in GA. With this proposed approach, which we refer to 
as an {\em improved GA}, we can easily 
estimate the performance of polar codes with SC decoding 
for long codes with low complexity. 
We then demonstrate simulation results with block lengths up to $N=2^{18}$ and 
show that the results match well with the estimated BLER 
results obtained by the analysis using the improved GA.
A similar but empirical modification that improves the performance of GA 
has been proposed by Ha~{\em et al.}~\cite{Ha2004}, 
and our analytical studies also serve as a justification of their approach.

Furthermore, we investigate a second construction 
based on the flipping probability of an LLR. 
Similar to the case of the GA method, this approach can 
estimate the BLER without simulation.
By tracing the flipping probability instead of its mean value,
it is still possible to construct codes with large length,
even though numerical results show its sub-optimality over the improved
GA as the code length increases.
Note that both the GA and LLR flipping approaches
are based on the
assumption that the LLR can be modeled as Gaussian.
Since this is an approximation,
its accuracy is evaluated by comparing the estimated performance
based on this approximation with that obtained by simulation.

The major contributions of this paper are summarized as follows:
\begin{itemize}
 \item 
The GA method should trace the evolution of the mean LLR value
through nonlinear transformation.
Based on the log-domain analysis of this transformation process,
a new algorithm with closed-form expressions 
that can 
trace the evolution
with improved accuracy 
over an AWGN channel 
is developed. 
\item 
How to design the key SNR parameter in the code construction algorithm
is discussed based on the estimated BLER. Using this measure,
the capacity-approaching behavior of polar codes with various code rates
is demonstrated
with low-complexity SC decoding.
\item The effectiveness of the proposed GA is investigated through comparison with the other 
construction alternatives of the same complexity order
by Monte-Carlo simulation with block-lengths as large as $N=2^{18}=262\,144$.
\end{itemize}
We note that the use of codeword lengths 
of $10^{5}$ or more
has been considered in several applications such as high-speed optical communications~\cite{Smith2012}, optical recording systems~\cite{Coene2001}, and flash memories~\cite{Qiu2018},
where product codes are often adopted due to practical constraints on encoding/decoding complexity 
associated with long codeword lengths. 
Therefore, we expect that polar codes could be an alternative solution in the future
as polar codes with SC decoding
can be implemented with low encoding/decoding complexity, 
provided that good polar codes with such long codeword lengths can be designed with reasonable complexity.

This paper is organized as follows. Section~\ref{sec:Pol} 
introduces 
notations 
associated with polar codes and decoding 
used throughout this work. The construction based on the conventional GA method
is
reviewed and
the proposed calculation approach is presented in 
Section~\ref{sec:GA}.
As an alternative and tractable
approach to the GA method, a construction based on LLR flipping probability
is described in Section~\ref{sec:alternative}.
Numerical examples are given in Section~\ref{sec:example},
where the capacity-approaching behavior based on the construction using
the improved GA is demonstrated, together with comprehensive comparison with the other approaches.
Finally, Section~\ref{sec:conc} concludes this work.

\section{Polar Codes and Successive Cancellation Decoding}
\label{sec:Pol}

\subsection{Polar Code}

Let ${\bf u} = (u_0, u_1, \ldots, u_{N-1}) \in {\mathbb F}_2^N$
denote a binary input vector of length $N$.
Let ${\bf G}_N \in {\mathbb F}_2^{N\times N}$ 
denote a generator matrix formulated as
\begin{align}
 {\bf G}_N = {\bf G}_{2}^{\otimes \, n}, \quad
{\bf G}_2 = \left( \begin{array}{cc}
1 & 0\\
1 & 1 \\
			  \end{array}\right),
\end{align}
where $n=\log_2 N$
and ${\bf A}^{\otimes\, n} = {\bf A} \otimes {\bf A}^{\otimes\,(n-1)}$ is the $n$th Kronecker power
of matrix ${\bf A}$
with ${\bf A}^{\otimes\, 0}  = (1)$~\cite{arikan09}.
The corresponding output 
${\bf x} = (x_0, x_1, \ldots, x_{N-1}) \in {\mathbb F}_2^N$
is expressed as
\begin{align}
 {\bf x} = {\bf u} \,{\bf B}_N {\bf G}_N,
\end{align}
where ${\bf B}_N \in {\mathbb F}_2^{N\times N}$ 
is the $N\times N$ bit reversal permutation matrix~\cite{arikan09}
which guarantees that SC decoding is performed in the bit reversal order
of the binary channel index.

We assume BPSK modulation over an AWGN channel, i.e., the transmitted symbol 
$s_i \in {\mathbb R}$, $i = 0, 1, \ldots, N-1$, is given by
\begin{align}
 s_i = \sqrt{E_s} (1 - 2 x_i ) ,
\label{eq:si}
\end{align}
with $E_s$ representing the symbol energy, and the received symbol $y_i \in {\mathbb R}$ is 
\begin{align}
y_i = s_i + z_i,
\label{eq:yi}
\end{align}
where $z_i \sim {\cal N}(0, N_0/2)$, i.e., $z_i$ is a real-valued Gaussian random variable
with zero mean and variance $N_0/2$. We denote the 
received signal vector of length $N=2^n$ as
${\bf y}_n = (y_0, y_1, \ldots, y_{2^n-1})$,
and its sub-vector of length $2^m$ starting from the index $i$
as ${\bf y}_m^{(i)} = (y_i, y_{i+1}, \ldots, y_{i+2^m-1})$.

\subsection{Successive Cancellation Decoding}

Based on 
SC decoding, 
given the length-$2^n$ received symbol observation vector ${\bf y}_n$,
each input bit $u_i$, $i=0,1, \ldots, N-1$, is decoded successively based on its
corresponding 
LLR.
We use the following short-hand notation for LLR of the $i$th input bit
$u_i$ as
\begin{align}
L_n^{(i)} 
({\bf y}_n)
= 
\begin{cases}
\log \frac{
p\left(
{\bf y}_n
 \,|\,
 u_{i} = 0
\right)
}
{
p\left(
{\bf y}_n
\,|\,
 u_{i} = 1 
\right)
}, & i = 0,\\
\log 
\frac{
p\left(
{\bf y}_n,\, \hat{\bf u}_{i-1} 
\,|\,
 u_{i} = 0 
\right)
}
{
p\left(
{\bf y}_n,\, \hat{\bf u}_{i-1} 
\,|\,
 u_{i} = 1 
\right)
}, & i = 1, 2, \ldots, 2^n-1,
\end{cases}
\label{eq:LLRdef}
\end{align}
where $p(\cdot | \cdot)$ 
denotes a conditional probability density function and
$\hat{\bf u}_{i-1} = (\hat{u}_0, \hat{u}_1, \ldots, \hat{u}_{i-1})$ 
is the vector containing
the estimated bits that have been already determined upon decoding of the $i$th bit.
(The subscript $n$ of $L_n^{(i)}$ reflects the 
fact that the LLR is calculated based on
the observation of $2^n$ received symbols in addition
to the previously estimated input bits.)
Note that for simplicity of notation, we will omit
the dependence of LLR 
on the previously estimated bits $\hat{\bf u}_{i-1}$ in what follows.

Because of the unique structure of polar codes,
the above LLRs 
can be recursively calculated 
for 
$k =1, 2, \ldots, n$ 
with $n=\log_2 N$ and 
$i = 0, 1, \ldots, 2^{k-1}-1$ 
as~\cite{arikan09,mori2009}
\begin{align}
L_{k}^{(2i)} 
\left(
{\bf y}_k
\right)
 & = 
L_{k-1}^{(i)}
\left(
{\bf y}_{k-1}^{(0)}
\right)
\boxplus
L_{k-1}^{(i)}  
\left(
{\bf y}_{k-1}^{(2^{k-1})}
\right)
\label{LN_upper0}
\\
L_{k}^{(2i+1)}  
\left(
{\bf y}_k
\right)
& = 
L_{k-1}^{(i)}
\left(
{\bf y}_{k-1}^{(2^{k-1})}
\right)
+ 
\left(
-1
\right)^{\hat{u}_{2i}} 
L_{k-1}^{(i)}  
\left(
{\bf y}_{k-1}^{(0)}
\right).
\label{LN_lower0}
\end{align}
Alternatively, by rewriting
$L_k^{(2i)} \triangleq L_{k}^{(2i)} \left({\bf y}_k\right)$,
$L_{{k-1}}^{(i)} \triangleq L_{k-1}^{(i)} \left({\bf y}_{k-1}^{(0)} \right)$,
and 
$L_{{k-1}}^{(i)'} \triangleq L_{k-1}^{(i)} \left( {\bf y}_{k-1}^{(2^{k-1})} \right)$
for simplicity, we have 
\begin{align}
L_{k}^{(2i)} 
 & = 
L_{k-1}^{(i)}
\boxplus
L_{k-1}^{(i)'}  
\label{LN_upper}
\\
L_{k}^{(2i+1)}  
& = 
L_{k-1}^{(i)'}
+ 
\left(
-1
\right)^{\hat{u}_{2i}} 
L_{k-1}^{(i)}  .
\label{LN_lower}
\end{align}
As noted in~\cite{mori2009},
\eqref{LN_upper} corresponds to the LLR calculation associated with
a check node, whereas~\eqref{LN_lower} corresponds to that with a variable node (or bit node)
in the Tanner graph. 
The operation $\boxplus$ of LLR is defined as~\cite{hage1996} 
\begin{align}
L_a \boxplus L_b =\log \frac{1+e^{L_a}e^{L_b}}{e^{L_a} + e^{L_b}}
= 2 \tanh^{-1} \left(
\tanh\left(\frac{L_a}{2}\right)
\tanh\left(\frac{L_b}{2}\right)
\right) .
\label{eq:oplus}
\end{align}
The recursive decoding procedure continues, starting from $k=n$ and until it reaches
$1$, 
where the last LLR
on the right hand side of \eqref{LN_upper0} and \eqref{LN_lower0}, i.e.,
$L_0^{(0)}(y_i)$, corresponds to the LLR of the channel bit $x_i$.
For an AWGN channel, we have 
\begin{align}
L_0^{(0)}(y_i)  = \log 
\frac{
p\left(
y_i
\,|\,
x_{i} = 0 
\right)
}
{
p\left(
y_i
\,|\,
x_{i} = 1 
\right)
}
=
4 \frac{\sqrt{E_s}}{N_0} y_i
, \qquad i = 0, 1, \ldots, N-1.
\label{eq:LLR1}
\end{align}

\subsection{Polar Code Construction}

For polar codes with rate $R = K/ N$,
$K$ channel indices out of $N$ total indices
are selected for information transmission.
Let ${\cal I} \subset \{0, 1, \ldots, N-1 \}$ denote the set 
of the channel indices selected to be information bits, i.e.,
the {\em information set}~\cite{arikan09},
with its cardinality
given by $|{\cal I}| =K$. The bits that are not 
in this set are called {\em frozen bits}
and fixed to known values (usually set as $0$).
Therefore, polar code construction is equivalent to the selection
of a set ${\cal I}$ that leads to a good block error rate~(BLER) 
performance.

In this work, we mostly focus on a Gaussian approximation~(GA) based design of polar codes.
This approach allows us to
construct good polar codes with much less complexity than those based on
precise calculation using density evolution. We demonstrate that the carefully designed
GA will also be able to predict the resulting BLER performance achieved by simulations
for a given code with almost negligible computational effort.

\section{Improved Gaussian Approximation} 
\label{sec:GA}

In this section, we first study the conventional GA approach, originally
proposed for design and analysis of LDPC codes, 
and its limitations when applied to the design of polar codes. 
Then, through careful design of associated metric calculations,
we show how to overcome these limitations by an improved GA.

\subsection{Gaussian Approximation}

Assume that the all-zero input sequence and thus the all-zero codeword is transmitted.
Then, from \eqref{eq:si} and \eqref{eq:yi},
$y_i \sim {\cal N}(\sqrt{E_s}, N_0/2 )$ for all $i$ and thus
from \eqref{eq:LLR1}
we observe that $L_0^{(0)} \triangleq L_0^{(0)}(y_i) \sim {\cal N}(\gamma_0, 2 \gamma_0)$
with $\gamma_0 = 4 E_s/N_0$.
When the value $E_s/N_0$ is used for 
construction of polar codes,
we refer to this value 
as the {\em design SNR}~\cite{trifonov2012,Vangala2015},
denoted by ${\rm SNR}_{\rm des}$.
Otherwise, $E_s/N_0$ corresponds to the channel SNR.
(Note that the design SNR in this work
is not given with respect to the conventional $E_b/N_0$ where $E_b$ is defined
as the energy {\em per information bit}. This is because
$E_b/N_0$ is a function of both $E_s/N_0$ and the code rate, and
thus is not convenient when comparing transmissions of 
codewords with the same symbol energy but different code rate.)
From \eqref{LN_upper}, \eqref{LN_lower},
and \eqref{eq:oplus} with the assumption that the previous bits 
are correctly estimated (i.e., $\hat{{\bf u}}_{i-1} ={\bf 0}_i$
for a given $i$, where ${\bf 0}_m$ denotes the zero vector of length $m$),
we have 
\begin{align}
\tanh\left(\frac{L_{{k}}^{(2i)}}{2}\right)
&
 =
\tanh\left( \frac{L_{{k-1}}^{(i)} }{2} \right)
\tanh\left(\frac{L_{{k-1}}^{(i)'}}{2}\right) ,
\label{eq:degrade}
\\
L_{{k}}^{(2i+1)}
& 
=
L_{{k-1}}^{(i)}+
L_{{k-1}}^{(i)'} ,
\label{eq:upgrade}
\end{align}
for 
$k = 1, 2,\ldots, n$, and $i = 0, 1, \ldots, 2^{k-1}-1$
with $n= \log_2 N$, 
where both
$L_{k-1}^{(i)}$ and $L_{k-1}^{(i)'}$ can be considered as
independent and identically distributed (i.i.d.) random variables
due to the structure of the polar codes with SC decoding.
Under the assumption that 
the LLRs $L^{(i)}_{k-1}$ and $L^{(i)'}_{k-1}$
follow a
${\cal N}(\gamma^{(i)}_{k-1}, 2 \gamma^{(i)}_{k-1})$ distribution, the only parameter
required to characterize its statistical distribution
is its mean. 
In fact, if this Gaussian assumption holds, 
since the two LLRs on the right hand side 
of  \eqref{eq:upgrade}
are statistically independent, it follows from
\eqref{eq:upgrade} 
that 
the output
$L^{(2i+1)}_{k}$ should precisely follow ${\cal N}(2 \gamma^{(i)}_{k-1}, 4 \gamma^{(i)}_{k-1})$, i.e.,
\begin{align}
 \gamma^{(2i+1)}_{k} = 2 \gamma^{(i)}_{k-1}.
\label{eq:gamma_upgrade}
\end{align}
On the other hand, the transformation~\eqref{eq:degrade}
changes the statistical distribution of the output 
$L^{(2i)}_{k}$ from Gaussian.
Nevertheless, Chung {\em et al.} applied statistical expectation to 
both sides of \eqref{eq:degrade}
to get~\cite{chung2001}
\begin{align}
E\left[
\tanh\left(\frac{L_{k}^{(2i)}}{2}\right)
\right]
&
 =
E\left[
\tanh\left( \frac{L_{k-1}^{(i)} }{2} \right)
\right]
E\left[
\tanh\left(\frac{L_{k-1}^{(i)'}}{2}\right) 
\right] \nonumber \\
& =
\left(
E\left[
\tanh\left( \frac{L_{k-1}^{(i)} }{2} \right)
\right]
\right)^2 .
\label{eq:degrade2}
\end{align}

Upon calculating 
\eqref{eq:degrade2},
let us define the monotonically increasing 
function $\psi(\gamma):(0,\infty) \to (0,1)$
with respect to the mean $\gamma$ of the random variable $L \sim {\cal N}(\gamma, 2\gamma)$ as
\begin{align}
 \psi(\gamma) & \triangleq E\left[
\tanh\left( \frac{L}{2} \right)
\right]
= 
\int_{-\infty}^{\infty}
\tanh\left( \frac{x}{2} \right) 
\frac{1}{\sqrt{4 \pi \gamma }}
e^{- \frac{\left( x - \gamma \right)^2}{4 \gamma}} dx.
\label{eq:psi}
\end{align}
Now, following~\cite{Trifonov2018},
let us introduce the short-hand notation for the compound function:
\begin{align}
 \Xi(\gamma) \triangleq 
 \psi^{-1}\left(
\psi^2\left(
\gamma
\right)
\right).
\label{eq:compound}
\end{align}
Then, we may rewrite \eqref{eq:degrade2} as
\begin{align}
\gamma^{(2i)}_{k} & =
\Xi
\left(
\gamma^{(i)}_{k-1}
\right) .
\label{eq:gamma2k}
\end{align}
Note that the 
compound 
function $\Xi(\gamma)
: (0,\infty) \to (0,\infty)$ is monotonically increasing.
However, 
since $\psi(\gamma)$ is strictly increasing in $\gamma$
with $0 < \psi(\gamma) < 1$, 
we have
$\psi^2(\gamma) < \psi(\gamma)$ and hence $\Xi(\gamma)=\psi^{-1}(\psi^2(\gamma)) < \gamma$.
Therefore, for $\gamma >0$, $\Xi(\gamma)$ is always a value that is strictly smaller than 
its input $\gamma$.

\begin{figure} [tbp]
\begin{center}
\includegraphics[width=.65\textwidth,clip]{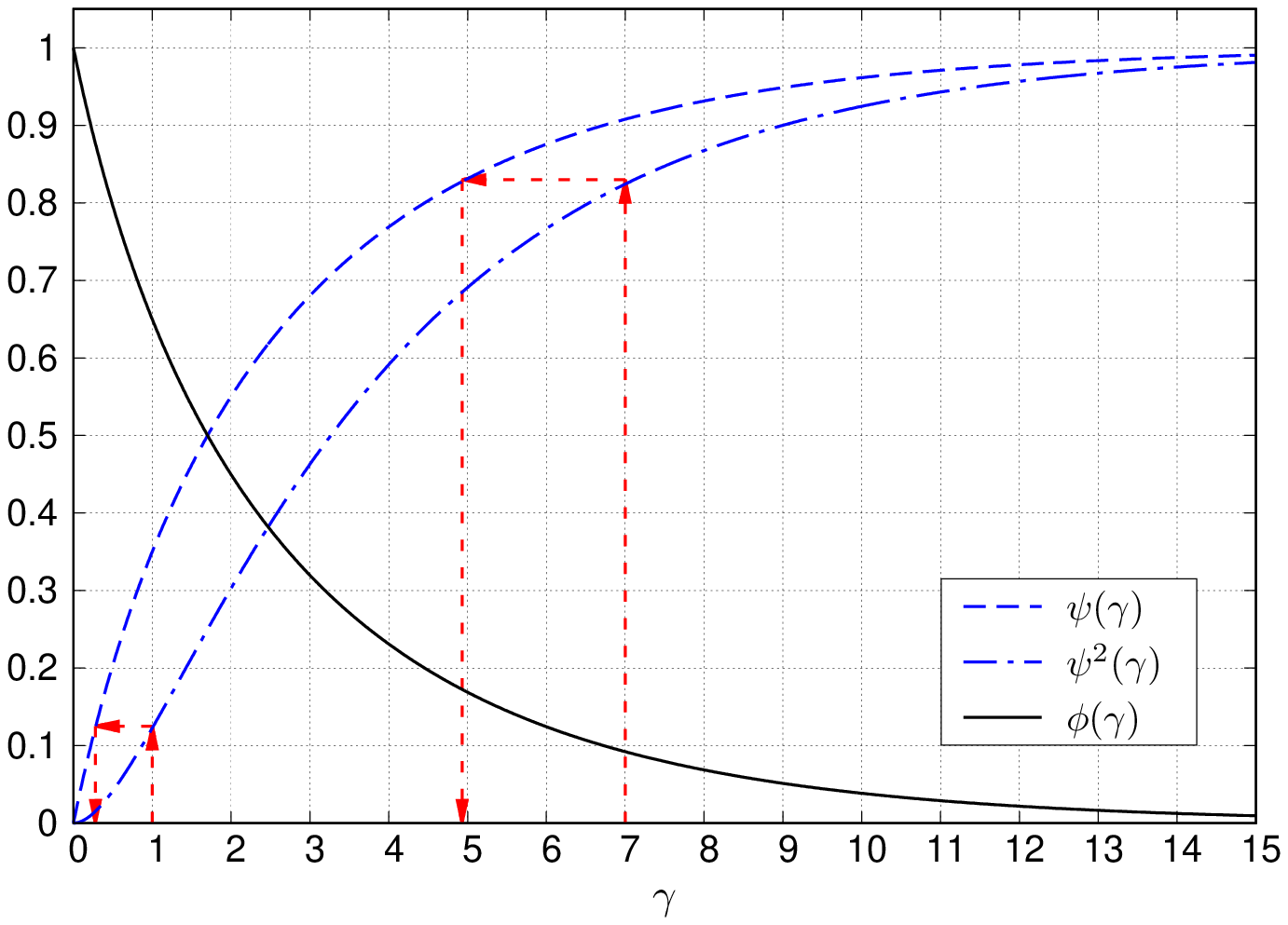}
\vspace{-.4cm}
 \caption{The mean LLR converting functions $\psi(\gamma), \phi(\gamma)$, and its associated function $\psi^2(\gamma)$. The trajectories demonstrate how the mean value is converted by the equation~\eqref{eq:compound}.}
\label{fig:mean_GA}
\vspace{-.2cm}
\end{center}
\vspace{-.4cm}
\end{figure}

Instead of directly dealing with $\psi(\gamma)$, 
Chung {\em et al.} introduce its complement function 
\begin{align}
\phi(\gamma) = 1 - \psi(\gamma).
\label{eq:phi}
\end{align}
In this case, $\phi(\gamma)$ is a monotonically decreasing function and approaches zero
as $\gamma \to \infty$, and
\eqref{eq:gamma2k} is equivalent to 
\begin{align}
\gamma^{(2i)}_{k} & =
 \phi^{-1}\left(
1-
\left(
1-
\phi
\left(
\gamma^{(i)}_{k-1}
\right)
\right)^2
\right).
\label{eq:gamma_degrade}
\end{align}
Fig.~\ref{fig:mean_GA} plots the associated functions $\psi(\gamma)$,
$\psi^2(\gamma)$, and $\phi(\gamma)$,
which are calculated by numerical integration.
The trajectories in the figure show how the mean value $\gamma$ is converted by
the transformation of \eqref{eq:compound},
and it is apparent that this
transformation always reduces the mean LLR value.

In~\cite{chung2001}, 
the following approximation for $\phi(\gamma)$ in~\eqref{eq:phi} 
was suggested:
\begin{align}
\phi (\gamma) & \approx
e^{a  \gamma^c + b} , \qquad \gamma \leq \Gamma_{\rm th},
\label{eq:chung_form}
\end{align}
where the constants were numerically determined as
$(a,b,c) = (-0.4527,0.0218,  0.86)$,
and the threshold was chosen as $\Gamma_{\rm th} \approx 10$.
For the case of $\gamma > \Gamma_{\rm th}$,
based on the relationship
\begin{align}
\sqrt{\frac{\pi}{\gamma}}
 e^{-\frac{\gamma}{4}}
\left(
1 - \frac{3}{\gamma}
\right)
<
\phi (\gamma) <
\sqrt{\frac{\pi}{\gamma}}
 e^{-\frac{\gamma}{4}}
\left(
1 + \frac{1}{7 \gamma}
\right),
\label{eq:chung_form2}
\end{align}
and since the upper and lower bounds converge as $\gamma \to \infty$,
it was suggested to approximate $\phi(\gamma)$ by the average of the upper and lower bounds in \eqref{eq:chung_form2}, i.e.,
\begin{align}
\phi (\gamma) \approx
\sqrt{\frac{\pi}{\gamma}}
 e^{-\frac{\gamma}{4}}
\left(
1 - \frac{10}{7 \gamma}
\right).
\label{eq:chung_form3}
\end{align}
For small $\gamma$, \eqref{eq:chung_form}
may not be  accurate as $\gamma \to 0$.
Therefore, 
even if \eqref{eq:chung_form} may be sufficient
for the moderate size of $N$ as confirmed in~\cite{wu2014,li_yuan2013},
the performance degradation 
may occur
for large code length $N$ 
due to numerical inaccuracy.

There are two numerical computation issues 
with the above GA approach.
The first issue is due to the approximation
associated with~\eqref{eq:chung_form} for a small value of $\gamma$,
which results in 
$\lim_{\gamma \to 0} \phi (\gamma )=e^b > 1 $.
As a consequence, when 
$\gamma$ approaches zero 
(i.e., once the mean LLR falls below a certain level),
the input to the inverse function $\phi^{-1}(\cdot)$
of \eqref{eq:gamma_degrade}
 fails to approach one and thus
 $\gamma$ cannot become zero, rendering
further polarization effects untraceable.
The second issue is due to the fact that the function $\phi(\gamma)$ 
takes a value in the interval $(0,1)$ and can thus 
approach arbitrarily close to zero as $\gamma$ becomes large.
For example, when $\gamma = 1000$,
which is a typical value in the case of polar codes with relatively
large code length, we observe that
 $\phi(\gamma) \approx 1.49 \times 10^{-110}$.
Since its inverse function $\phi^{-1}(\cdot)$ cannot be expressed in a 
closed form, one should resort to numerical approaches such as 
the bisection method for improving the 
numerical accuracy.
As $\gamma$ becomes large
(i.e., 
once the mean LLR exceeds a certain level), however,
numerically computing the inverse function 
fails to return an accurate value even with the bisection method, and thus again makes
further polarization untraceable\footnote{%
As we shall discuss in Remark~\ref{rem:asympto} of Section~\ref{subsec:prop},
once the mean LLR of given channels exceeds some large value at a given stage, 
it may diminish slowly even if it experiences $\boxplus$ operations in later stages. 
Therefore, one may stop further tracing for these channels as they are
most likely to be selected as information bits.
}.
These issues can be solved by the proposed log-domain
calculation technique of the function $\phi(\gamma)$
described in what follows.

\begin{remark}
The fact that the approximation suggested by Chung {\em et al.} 
should result in $\lim_{\gamma \to 0 } \phi(\gamma) >1$
was identified by Ha {\em et al.}~\cite{Ha2004},
in conjunction with 
designing {\em punctured} LDPC codes.
(This fact has been also pointed out in~\cite{dai2017access} in the framework of 
polar code design with long codewords.)
Therefore, in addition to~\eqref{eq:chung_form},
the following correction has been proposed:
\begin{align}
\phi(\gamma) 
\approx e^{\alpha \gamma + \beta \gamma^2}, \qquad {\rm for}\quad 0 <  \gamma < \Gamma'_{\rm th},
\label{eq:Ha}
\end{align}
where $\alpha=-0.4856$,  $\beta = 0.0564$, and $\Gamma'_{\rm th} = 0.867861$.
This approach resolves the issue associated with the numerical 
inaccuracy of the original GA as $\gamma \to 0$. Similar expressions have been developed
in~\cite{dai2017access}.
\end{remark}

\subsection{Log-Domain Expression for $\phi(\gamma)$}

Recall that the mean LLR of polar codes at the code construction
design stage should range from very small values to very large values,
and the metric calculation associated with GA should support this exponential range.
For example,
if the design SNR is 1\,(i.e., 0\,dB),
then the initial mean LLR value should be $\gamma_0 = 4$.
For a polar code with $N= 2^n$, since
each operation of \eqref{eq:upgrade} doubles the LLR value as shown in \eqref{eq:gamma_upgrade},
the largest value (after $n$ stages) is $L_{\max} = 4 \times 2^n$. 
On the other hand,
the amount of change from $\gamma^{(i)}_{k-1}$ to $\gamma^{(2i)}_{k}$ 
according to \eqref{eq:degrade} depends on 
its input value, and as is observed from Fig.~\ref{fig:mean_GA},
its ratio $\gamma^{(2i)}_{k}/\gamma^{(i)}_{k-1}$  rapidly becomes smaller
as $\gamma^{(i)}_{k-1}$ decreases and the ratio becomes much less than $1/2$
with small enough $\gamma^{(i)}_{k-1}$. 
Therefore, in order to track the mean values by GA accurately,
$\gamma^{(2i)}_k$ must be precisely calculated even if it becomes extremely small as discussed
in the previous subsection.

To cope with this issue, we define the logarithmic domain of $\phi(\gamma)$ as
\begin{align}
\xi(\gamma) \triangleq \log \phi(\gamma) = \log( 1 - \psi(\gamma))  
\label{eq:xi_def}
\end{align}
and based on this function we 
attempt to trace the mean LLR value.
Note that
$\xi(\gamma):(0,\infty) \to (-\infty, 0)$, and $\xi(\gamma)$ is monotonically 
decreasing
with $\gamma$. 
Two major advantages of using \eqref{eq:xi_def} are that
i) as $\gamma \rightarrow 0$, $\xi(\gamma) \approx - \psi(\gamma)$ since $\psi(\gamma) \rightarrow 0$ as well, and thus
$\xi(\gamma)$ can accurately trace $\gamma$ in this regime as well,
and ii) for $\gamma \rightarrow \infty$, then $\psi(\gamma) \rightarrow 1$, and it will be shown that 
$\xi(\gamma)$ is, to first order, linear in $\gamma$, and thus $\gamma$ can again be accurately traced in this regime.
By comparison, when $\gamma$ is large, $\psi(\gamma) \approx 1$ (see Fig.~\ref{fig:mean_GA}), and $\psi(\gamma)$ would need to be computed with very high precision to obtain $\gamma$ from its inverse as even a slight error in $\psi(\gamma)$ can result in an incorrect inverse $\gamma$.
Therefore, we first develop 
the asymptotic forms of $\xi(\gamma)$ for low and high values of $\gamma$.

\subsubsection{Asymptotic Form of $\xi(\gamma)$ for Small Values of $\gamma$}

We first note that the following lemma holds for the function $\psi(\gamma)$:%
\begin{lemma}\label{th:lemma}
For $\psi(\gamma)$ defined in \eqref{eq:psi}, we have
\begin{align}
 \psi(\gamma)
& = \frac{1}{2} \gamma - \frac{1}{4}\gamma^2 + \frac{5}{24} \gamma^3 + O\left( \gamma^4 \right) 
\qquad {\rm as} \quad \gamma \to 0.
\label{eq:th01}
\end{align}
\end{lemma}
The proof is given in Appendix~\ref{app:proof}.

From Lemma~\ref{th:lemma}, 
the following theorem is immediate by 
a Maclaurin series expansion of $\xi(\gamma)$:
\begin{theorem}\label{th:theorem}
For $\xi(\gamma)$ defined in \eqref{eq:xi_def}, we have
\begin{align}
\xi( \gamma)  = - \frac{1}{2}\gamma +\frac{1}{8}\gamma^2 
-\frac{1}{8}\gamma^3 + 
O(\gamma^4)
\qquad {\rm as} \quad \gamma \to 0.
\label{eq:th02}
\end{align}
\end{theorem}

\subsubsection{Asymptotic Form of $\xi(\gamma)$ for Large Values of $\gamma$}
In this case, we may make use of the following expression:
\begin{theorem}\label{th:theorem2}
For $\xi(\gamma)$ defined in \eqref{eq:xi_def}, we may express
\begin{align}
\xi(\gamma) 
=
-\frac{\gamma}{4}
+
\frac{1}{2} \log \pi
- \frac{1}{2} \log \gamma
+\log\left(
1 - \frac{\pi^2}{4 \gamma} + 
O\left(
\frac{1}{\gamma^2}
\right)
\right) 
\qquad {\rm as} \quad \gamma \to \infty.
\label{eq:th03}
\end{align}
\end{theorem}

\begin{IEEEproof}
By applying the identity $\tanh(x/2) = 1-2/(e^x + 1)$
to \eqref{eq:psi}, 
we may 
rewrite \eqref{eq:phi}
directly through series expansion as 
\begin{align}
\phi(\gamma)& =
\frac{1}{\sqrt{\pi \gamma }}
\int_{-\infty}^{\infty}\frac{1}{e^x + 1} 
e^{- \frac{\left( x - \gamma \right)^2}{4 \gamma}} dx
\nonumber \\
& =
\frac{e^{-\frac{\gamma}{4}}}{\sqrt{\pi \gamma }}
\int_{-\infty}^{\infty}\frac{e^{\frac{x}{2}}}{e^x + 1} 
e^{- \frac{x^2}{4 \gamma}} dx \nonumber \\
& =
\frac{e^{-\frac{\gamma}{4}}}{\sqrt{\pi \gamma }}
\int_{-\infty}^{\infty}\frac{e^{\frac{x}{2}}}{e^x + 1} 
\sum_{k=0}^{\infty} \frac{1}{k!}
\left(
- \frac{x^2}{4 \gamma}
\right)^k
 dx
\nonumber \\
&= 
\frac{e^{-\frac{\gamma}{4}}}{\sqrt{\pi \gamma }}
\sum_{k=0}^{\infty} \frac{(-1)^k}{k! (4 \gamma)^k}
\int_{-\infty}^{\infty}
x^{2k}
\frac{e^{\frac{x}{2}}}{e^x + 1} 
dx \nonumber \\
& =
\frac{e^{-\frac{\gamma}{4}}}{\sqrt{\pi \gamma }}
\sum_{k=0}^{\infty} \frac{(-1)^k}{(16 \gamma)^k}
\frac{(2k)!}{k!}
\left\{
\zeta\left(
 2k + 1, \frac{1}{4}
\right)
-
\zeta\left(
 2k + 1, \frac{3}{4}
\right)
\right\}\nonumber \\
& =
\frac{e^{-\frac{\gamma}{4}}}{\sqrt{\gamma }}
\sqrt{\pi}
\left(
1 - \frac{\pi^2}{4 \gamma} + \frac{5\pi^4}{32 \gamma^2}
- \frac{61 \pi^6}{384 \gamma^3} + \cdots
\right),
\label{eq:phi4}
\end{align}
where $\zeta( m, q)$ is the generalized Riemann zeta function
defined as
\begin{align}
\zeta(m,q) = \sum_{l=0}^{\infty}\frac{1}{(l + q)^m}.
\end{align}
Taking logarithm of both sides of \eqref{eq:phi4}
leads to~\eqref{eq:th03}.
\end{IEEEproof}

Note that the asymptotic case of
Theorem~\ref{th:theorem2}
agrees with the bounds developed in~\cite{chung2001}, i.e., \eqref{eq:chung_form2}.

\subsubsection{Closed-Form Approximation}

Given the above observations of  $\xi(\gamma)$ in term of limiting behavior,
the remaining issue is how to describe its closed form approximation $\hat{\xi}(\gamma)$ 
for each value of $\gamma$.
We propose the following expression:
\begin{subnumcases}{\label{eq:proposed_xi}  \hat{\xi}(\gamma)  =}
-\frac{1}{2} \gamma + \frac{1}{8} \gamma^2 - \frac{1}{8}\gamma^3,  &  $\gamma \leq \Gamma_0$,
\label{eq:case_low}\\
 a_0 + a_1 \gamma 
+ a_2 \gamma^2, & $\Gamma_{0} < \gamma \leq \Gamma_{1}$,
\label{eq:Ha_formula1}\\
 a \gamma^{c}
+ b, & $\Gamma_{1} < \gamma < \Gamma_{2}$,
\label{eq:Chung_formula1}\\
-\frac{\gamma}{4}
+
\frac{1}{2} \log \pi
- \frac{1}{2} \log \gamma
+\log\left(
1 -
 \frac{\pi^2}{4\gamma}
+ \frac{\kappa_0}{\gamma^2}
\right),
& $\Gamma_{2} \leq \gamma$,
\label{eq:large_gamma}
\end{subnumcases}
where 
the parameters in~\eqref{eq:Ha_formula1}
are chosen as $(a_0, a_1, a_2) = (-0.002706, -0.476711, 0.0512)$,
those in~\eqref{eq:Chung_formula1}
are adopted as 
$(a,b,c) = (-0.4527,0.0218,  0.86)$ 
following Chung {\em et al.}, 
and 
$\kappa_0 = 8.554$, with the thresholds
chosen as $\Gamma_0 = 0.2$, $\Gamma_1=0.7$, and $\Gamma_2 = 10$.

Each expression in \eqref{eq:proposed_xi} is chosen as follows:
When $\gamma$ is close to zero, 
\eqref{eq:case_low}
results from Theorem~\ref{th:theorem}, whereas
the expression~\eqref{eq:Chung_formula1} 
is inherited from the work of Chung {\em et al.}~\cite{chung2001}, i.e., 
\eqref{eq:chung_form}.
For a given pair of thresholds $\Gamma_0$ and $\Gamma_1$, 
the intermediate expression~\eqref{eq:Ha_formula1} is obtained by curve fitting,
where $a_2$ is optimized  through an exhaustive search such that it minimizes the squared 
error from the function calculated by numerical integration, provided that 
the piece-wise continuity should be guaranteed
at the boundaries $\Gamma_0$ and $\Gamma_1$ by selection of the parameters $a_0$ and $a_1$.
Finally,
for large values of $\gamma$, \eqref{eq:large_gamma} is adopted
with reference to \eqref{eq:th03},
where we have introduced a positive constant 
$\kappa_0$ as
a fitting parameter
that minimizes the squared error in the range $(\Gamma_2 ,\Gamma_2 + 2)$.
It is clear that the impact of $\kappa_0$ becomes negligible 
as $\gamma$ increases.

\begin{figure} [tbp]
\begin{center}
\includegraphics[width=.65\textwidth,clip]{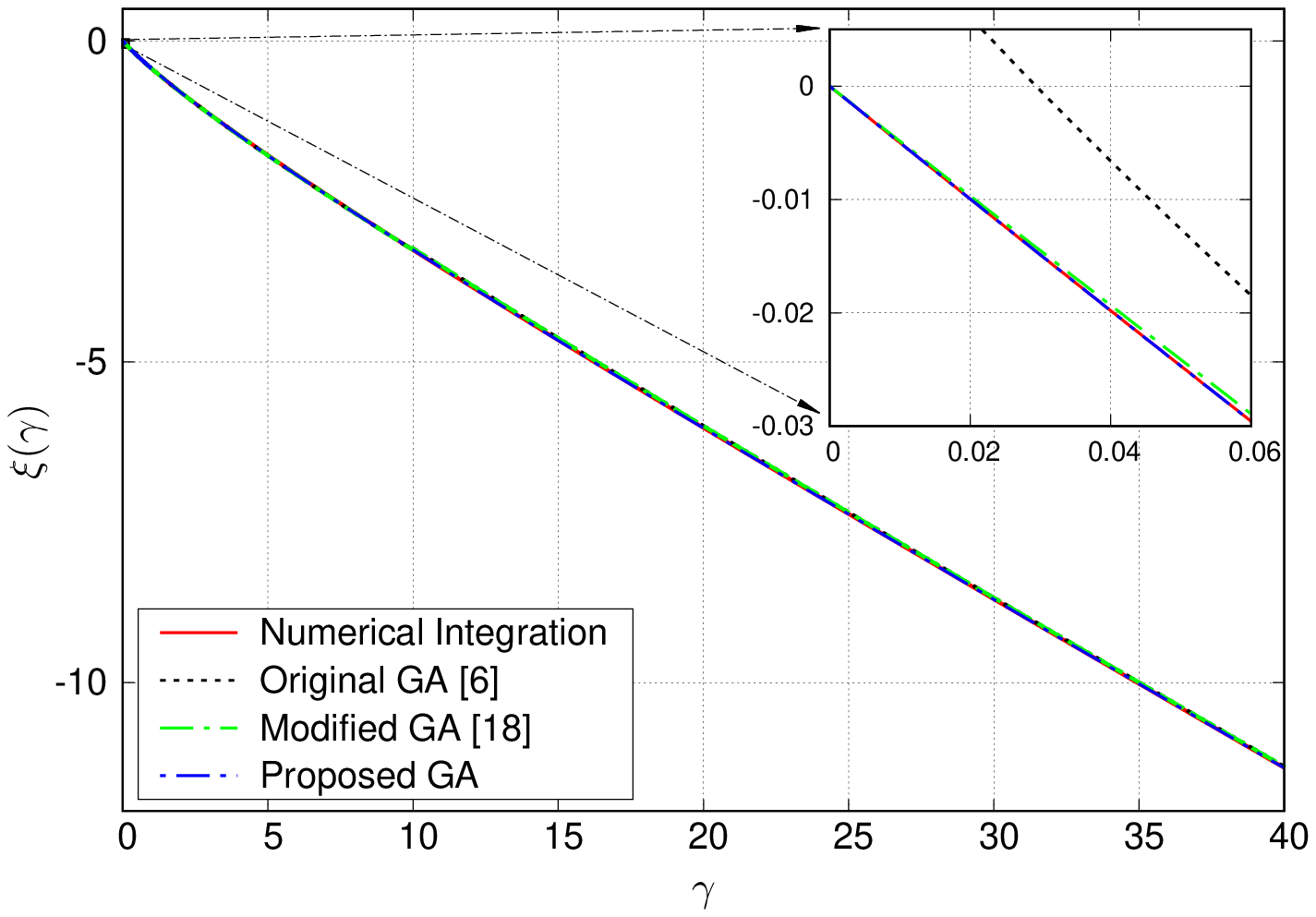}
\vspace{-.4cm}
 \caption{
Comparison of the function $\xi(\gamma)$ calculated through numerical
integration and its piece-wise approximate form $\hat{\xi}(\gamma)$
through~\eqref{eq:case_low}--\eqref{eq:large_gamma}.
Also plotted are the corresponding curves based on the conventional GA
by Chung {\em et al.}~\cite{chung2001} and its modified version
by Ha {\em et al.}~\cite{Ha2004}.
The area around the origin is enlarged as a reference.}
\label{fig:compare_curves}
\vspace{-.2cm}
\end{center}
\vspace{-.6cm}
\end{figure}

\begin{remark}
Note that the modified approximation formula proposed by Ha {\em et al.}~\cite{Ha2004},
i.e., \eqref{eq:Ha}, has strong similarity with the theoretical limit
of~\eqref{eq:case_low} if it is converted to the log domain.
Therefore, it may be capable of capturing the polarization effect in the case of $\gamma \to 0$,
similar to the proposed approach based on the log-domain analysis.
The corresponding approximate function~\eqref{eq:proposed_xi} 
together with those calculated using 
the conventional GA of~\eqref{eq:chung_form}
as well as its modification of~\eqref{eq:Ha} by Ha {\em et al.} 
are compared in Fig.~\ref{fig:compare_curves}.
Even though all the approximate curves exhibit good agreement with that based on the numerical
calculation, they reveal some difference in the proximity of $\gamma = 0$.
The conventional expression~\eqref{eq:chung_form} exceeds $0$, whereas
a slight difference is observed with the modification in~\eqref{eq:Ha}.
As expected,
the proposed 
expression yields no discernible difference
from the curve calculated by numerical integration for $\gamma$ near $0$.
\end{remark}

\subsection{Proposed Log-Domain GA Algorithm}
\label{subsec:prop}

Given the expression of $\xi(\gamma)$,
the mean LLR value associated with \eqref{eq:gamma_degrade} 
can be rewritten with respect to $\xi(\gamma)$ as
\begin{align}
\gamma^{(2i)}_{k}
& = \xi^{-1}\left(
\xi\left(
\gamma_{k-1}^{(i)}
\right)
+ \log\left(
2 - e^{\xi\left(
\gamma_{k-1}^{(i)}
\right)}
\right)
\right),
\label{eq:gamma_propose}
\end{align}
whereas
$\gamma^{(2i+1)}_{k}$ is revised according to \eqref{eq:gamma_upgrade}.

The inverse function of $z=\hat{\xi}(\gamma)$,
i.e., 
$\hat{\xi}^{-1}(z):(-\infty, 0) \to (0,\infty)$ 
can be obtained by solving the corresponding 
equations as
\begin{subnumcases}{\label{eq:proposed_inv_xi}  \hat{\xi}^{-1}(z)  =}
- 2 z + z^2 + z^3, & $z \geq Z_0$,
\label{eq:hat_xi_low}\\
\frac{-a_1 - \sqrt{a_1^2 - 4 a_2 (a_0-z)}}{2 a_2}, & $ Z_1 \leq z < Z_0$,
\label{eq:hat_xi_middle}\\
\left(
\frac{z - b}{a}
\right)^{\frac{1}{c}}, & $ Z_2 < z < Z_1$,
\end{subnumcases}
where the $Z_i$'s are the thresholds 
of the two functions, i.e., $Z_i =\hat{\xi} \left(\Gamma_i\right)$.
Note that \eqref{eq:hat_xi_low} is obtained 
as an approximation,
through series expansion, of the inverse function
corresponding to~\eqref{eq:case_low}.
In the case of $z \leq Z_2$, the inverse function corresponding to
\eqref{eq:large_gamma}  may not be expressed in closed form. However,
it can be easily calculated with high accuracy by using the bisection method.

  \begin{algorithm}[t]
   \caption{Channel Polarization with Improved GA}
   \begin{algorithmic}[1]
    \REQUIRE $n = \log_2 N$, $\alpha = {\rm SNR}_{\rm des}$
    \ENSURE $\gamma[0], \gamma[1], \ldots, \gamma[N-1]$ as $\gamma_{n}^{(0)}, 
  \gamma_{n}^{(1)}, 
  \ldots \gamma_{n}^{(N-1)}$ 
    \STATE $\gamma[0] = 4 \alpha$
    \FOR{$i=1$ : $n$}
    \STATE $J = 2^{i}$
    \FOR{$j=0$ : $J/2-1$}
    \STATE $u = \gamma[j]$ 
    \IF{$u \leq \Gamma_0$} 
    \STATE Update $\gamma[j]$ with $u$ as input using \eqref{eq:direct_Xi}
    \ELSE
    \STATE $z = \xi\left( u \right)$ using \eqref{eq:proposed_xi}
    \STATE $\gamma[j] = \xi^{-1}\left( z + \log\left(2 - e^{z} \right) \right)$ 
using \eqref{eq:proposed_inv_xi}
    \ENDIF
    \STATE $\gamma[j+J/2] = 2 u$
    \ENDFOR
    \ENDFOR
    \RETURN $\gamma[0], \gamma[1], \ldots, \gamma[N-1]$ 
   \end{algorithmic}
   \label{alg:proposedGA}
  \end{algorithm}

Furthermore, in the case of $\gamma \leq \Gamma_0$, 
by substituting $\hat{\xi}(\gamma)$ of \eqref{eq:case_low} 
into 
$\xi\left(\gamma_{k-1}^{(i)}\right)$ of \eqref{eq:gamma_propose},
and then substituting 
$\hat{\xi}\left(
\gamma
\right)
+ \log\left(
2 - e^{\hat{\xi}\left(
\gamma
\right)}\right)$ into $z$ of \eqref{eq:hat_xi_low} 
with
series expansion and collecting the dominant terms up to the fourth degree, 
one may obtain the following expression for $\Xi(\gamma)$
in \eqref{eq:compound}:
\begin{align}
\Xi(\gamma)& \approx 
\frac{1}{2}
\gamma^2
-
\frac{1}{2}
\gamma^3
+ \frac{2}{3} 
\gamma^4,
\qquad  \gamma \leq \Gamma_0.
\label{eq:direct_Xi}  
\end{align}

The improved GA algorithm outlined above 
is summarized in Algorithm~\ref{alg:proposedGA}.

\begin{remark}
\label{rem:asympto}
It would be of interest to investigate how $\Xi(\gamma)$ may behave as $\gamma \to \infty$.
Noticing that the logarithmic term 
grows substantially slower than the linear term,
one may ignore the contribution of 
the third and fourth terms in \eqref{eq:large_gamma}
and define the corresponding inverse as $\hat{\xi}^{-1}(z) = -4 z + 2 \log \pi$.
Then, by the same process that has lead to~\eqref{eq:direct_Xi},
we obtain
\begin{align}
\Xi(\gamma)  \approx \gamma - 4 \log 2, \qquad \gamma \to \infty.
\label{eq:gamma_asympto}
\end{align}
The above approximation indicates that,
when the input mean LLR value $\gamma$ is large, the $\boxplus$ calculation of
two independent and equivalent channels, i.e.,~\eqref{LN_upper},
will reduce the output mean LLR value approximately by 
$4\log 2 \approx 2.7726$, whereas it always doubles $\gamma$
when \eqref{LN_lower} is applied.
In other words, once $\gamma$ reaches a large enough value where
the approximation~\eqref{eq:gamma_asympto} holds,
then the rate at which $\gamma$
diminishes due to $\Xi(\gamma)$ is much less significant
and thus
these channels could be selected as information bits even without further 
high-accuracy tracing.
\end{remark}

\begin{figure} [tbp]
\begin{center}
\includegraphics[width=.65\textwidth,clip]{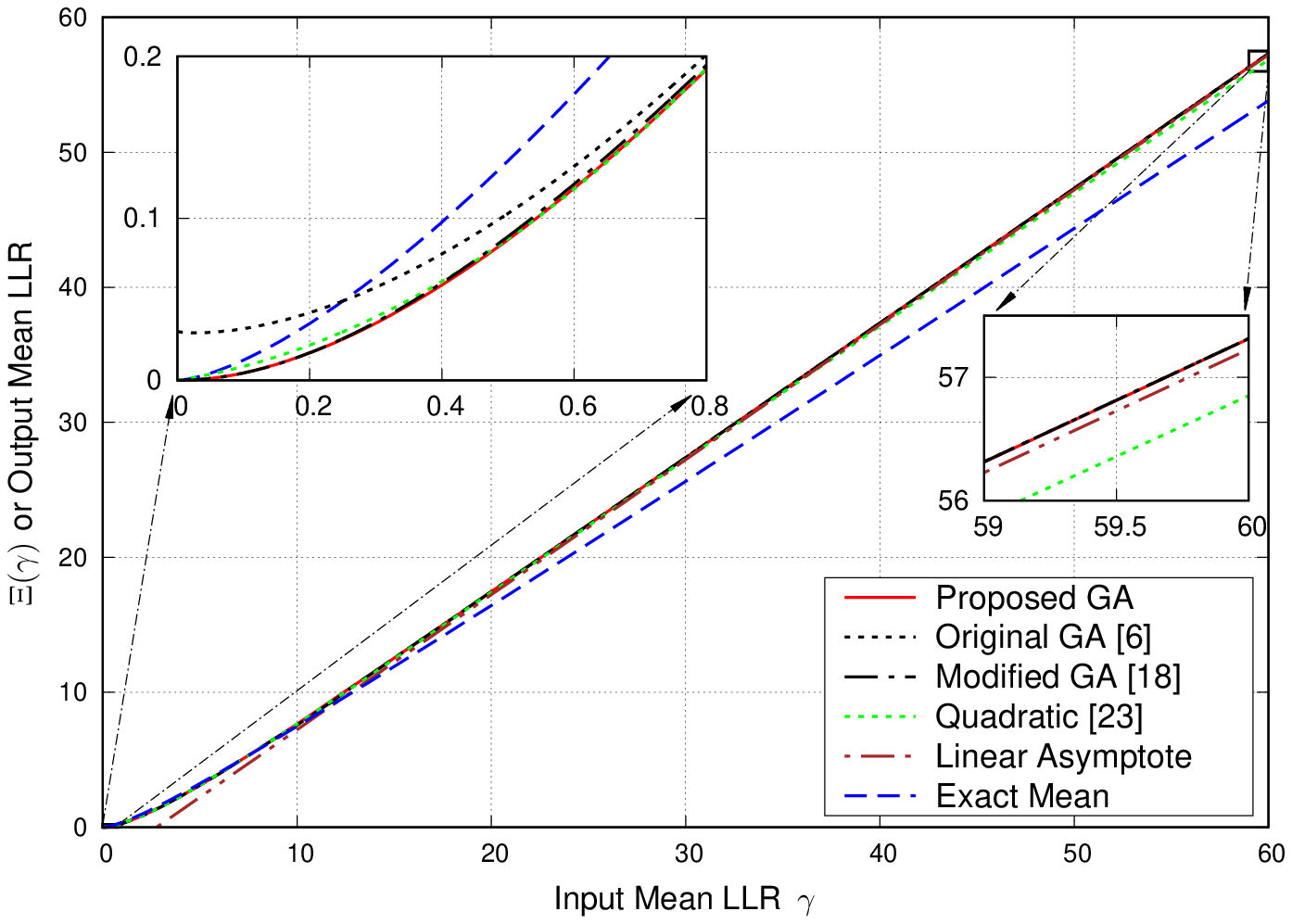}
\vspace{-.4cm}
 \caption{Comparison of the function $\Xi(\gamma)$
based on 
Algorithm~\ref{alg:proposedGA} as well as the other approaches, i.e.,
the original formula by Chung {\em et al.}~\cite{chung2001},
its modification by Ha {\em et al.}~\cite{Ha2004}, and 
a piece-wise quadratic function approximation 
by Trifonov~\cite{Trifonov2018}, together with the asymptotic form~\eqref{eq:gamma_asympto}.
Also shown as ``Exact Mean'' is the curve representing
the theoretical output mean LLR value corresponding to $\boxplus$ operation 
with Gaussian input ${\cal N}(\gamma, 2\gamma)$ based on \eqref{eq:direct}.
The two edge areas are enlarged as a reference.
}
\label{fig:mean_LLR}
\vspace{-.2cm}
\end{center}
\vspace{-.8cm}
\end{figure}

\begin{remark}
Fig.~\ref{fig:mean_LLR} shows 
the relationship between the input and output mean LLR values,
i.e., the function $\Xi(\gamma)$, 
obtained through the improved GA as well as those calculated by the other schemes discussed above. 
Also plotted are the linear asymptote given by~\eqref{eq:gamma_asympto}
and the mean LLR value directly calculated without assuming the GA model
for later reference (to be discussed in 
Remark~\ref{remark:GA} at the end of this section).
We observe 
some noticeable discrepancy between the improved GA and the original GA
as $\gamma$ approaches zero. However, the improved approximation by Ha {\em et al.}
resembles that of the improved GA developed here.
Also observed is the fact that the simple approximation~\eqref{eq:gamma_asympto} becomes tighter as
$\gamma$  increases.
We note that instead of dealing with the function $\phi(\gamma)$ and its inverse,
Trifonov~\cite{Trifonov2018}
introduces
a direct approximation of $\Xi(\gamma)$ 
by a piece-wise quadratic function, which is also plotted in Fig.~\ref{fig:mean_LLR}
for comparison. This approach may significantly simplify the design 
of polar codes and 
it turns out to be effective for constructing a wide range of polar codes
with moderate lengths.
However, we observe that there is some discrepancy especially
in the low value of $\gamma$
as our analysis suggests that their asymptotic form should 
follow~\eqref{eq:direct_Xi}, i.e., the quadratic approximation may have some non-negligible numerical gap.
This will affect the achievable performance
of the code designed
for long code lengths and low code rate
as will be demonstrated in Section~\ref{sec:example}.
\end{remark}

\subsection{Block Error Rate Estimation}

As discussed in \cite{mori2009,tal2013,wu2014}, 
once the distribution
of LLRs is determined by density evolution or the GA method,
its block error rate~(BLER) based on SC decoding
can also be estimated.
Assuming that the all zero codeword is transmitted, the bit error probability 
of the $i$th bit, provided that all the previous bits are correctly decoded,
 is given by
\begin{align}
P_{b, i} = \Pr\left( \hat{u}_i \neq 0 \left| \hat{u}_{0} = \cdots = \hat{u}_{i-1} = 0 \right. \right)
= \Pr\left( L_{n}^{(i)} < 0 \right),
\end{align}
and under the assumption that $L_{n}^{(i)} \sim {\cal N}(\gamma_{n}^{(i)},2\gamma_{n}^{(i)})$,
it follows that
\begin{align}
P_{b, i} =
 Q\left(
\sqrt{
\frac{\gamma_n^{(i)}}{2}}\right),
\label{eq:BER_bit_channel}
\end{align}
where the $Q$-function $Q(x):[0,\infty) \to \left(0, \frac{1}{2}\right]$
is defined as
$Q(x) = 
\frac{1}{2}{\rm{erfc}}\left(
\frac{x}{\sqrt{2}}
\right)=
\frac{1}{\sqrt{2\pi}} \int_{x}^{\infty} e^{-\frac{t^2}{2}} dt$.
The BLER is 
often approximated
 as
\begin{align}
P_{BL} & \approx 1 -
\prod_{i=0}^{N-1} 
\left(
1 -
P_{b, i} 
\right).
\end{align}

In the process of polar code design, let 
$\hat{\gamma}_n^{(i)}(\alpha)$, $i \in \{0, 1, \cdots, N-1\}$,
denote the estimated output mean LLR obtained by some GA algorithm
with $\alpha$ as its initial input SNR. 
If we select the set of $K$ channel indices with largest 
outputs
as the information set ${\cal I}$,
the minimum BLER with a given rate $R = K/N$
can be achieved. 
We define
the resulting {\em minimum estimated BLER}
for a given design SNR 
as
\begin{align}
\hat{P}_{BL,\min} (R, {\rm SNR}_{\rm des}) & \triangleq
\min_{{\cal I}: \,
\left|
{\cal I}\right| = R N}
\left\{
1 -
\prod_{i \in {\cal I}}
\left(
1-
 Q\left(
\sqrt{
\frac{\hat{\gamma}_n^{(i)}
\left(
{\rm SNR}_{\rm des}
\right) 
}{2}}
\right)
\right)
\right\}
.
\label{eq:BLER}
\end{align}
For a given pair of design SNR and code rate, the above BLER can be calculated 
by sorting ${\hat{\gamma}_n^{(i)}}({\rm SNR}_{\rm des})$ 
in decreasing order (after applying a specific GA algorithm)
and selecting only the largest $\left|{\cal I}\right|$ elements
from the entire index set $\{0,1,\cdots, N-1\}$.
Note that the above estimated performance corresponds to the scenario
where a code is designed for a specific SNR value and therefore
this performance is achieved by a {\em family} of polar codes with SC decoding, 
possibly employing a different information set at each SNR.

The above relationship can also be utilized for
evaluating the estimated performance of 
{\em a given designed polar code}. Let ${\cal I}^*$ denote the information set
designed through the above process, i.e., 
the set ${\cal I}$ that yields the minimum estimated BLER 
evaluated at a given design SNR well below a target BLER
according to~\eqref{eq:BLER}.
The resulting estimated BLER of the specific polar code defined by ${\cal I}^*$,
evaluated at a given {\em channel SNR}, 
may be defined as
\begin{align}
\hat{P}_{BL} \left(
{\cal I}^*, 
E_s/N_0
\right) & \triangleq
1 -
\prod_{i \in {\cal I}^*}
\left(
1-
 Q\left(
\sqrt{
\frac{\hat{\gamma}_n^{(i)}
\left(
E_s/N_0
\right) 
}{2}}
\right)
\right)
.
\label{eq:BLER2}
\end{align}
It is easy to see that~\eqref{eq:BLER} 
serves as a lower bound on~\eqref{eq:BLER2}. 
Moreover, if ${\cal I}^*$ of~\eqref{eq:BLER2} is designed 
with ${\rm SNR}_{\rm des}$ through~\eqref{eq:BLER},
then both BLER values should agree
at the channel SNR where the equality $E_s/N_0={\rm SNR}_{\rm des}$ holds.

\begin{remark}
\label{remark:GA}
We end this section by pointing out an intriguing fact of GA.
Even though  the initial LLR of bit channels
$L_{0}^{(0)}$ is Gaussian distributed,
there is no guarantee that $L_{k}^{(2i)}$ is Gaussian
since
the output after $\boxplus$ operation, i.e.,~\eqref{LN_upper}, is not strictly Gaussian. 
However, the notable aspect associated with the use of~\eqref{eq:degrade2}
is that even if $L_{k}^{(2i)}$ is not Gaussian,
the mean $\gamma^{(2i)}_k$ 
is chosen such that 
$L_{k}^{(2i)}$ is {\em approximated} by a Gaussian with mean $\gamma^{(2i)}_k$ and variance $2\gamma^{(2i)}_k$.
Therefore, the computed $\gamma^{(2i)}_k$ will be different from
a direct calculation of the LLR mean:
\begin{align}
E\left[
L_{k}^{(2i)} 
\right] & = 2 E\left[
\tanh^{-1}
\left\{
\tanh\left( \frac{L_{k-1}^{(i)} }{2} \right)
\tanh\left(\frac{L_{k-1}^{(i)'}}{2}\right) 
\right\}
\right].
\label{eq:direct}
\end{align}
In {\sc Appendix}~\ref{app:meanLLR}, 
the above mean value is theoretically derived,
and the results are compared in Fig.~\ref{fig:mean_LLR} 
with the mean LLR obtained by the (improved) GA for a given input mean LLR.
From the figure, we observe that the behavior of the two functions are significantly
different, especially as the mean input LLR increases. Due to this difference,
the polar code design using the mean LLR directly calculated through \eqref{eq:direct} 
may not lead to the same result
as the GA approach through \eqref{eq:degrade2}, and 
their performance is in general 
inferior to those designed by GA.
\end{remark}

\section{An Alternative Construction Based on LLR Flipping Probability}
\label{sec:alternative}

In the previous section, we discussed numerical issues associated 
with mean LLR calculation in the GA method,
and how to cope with them.
However, the non-linearity of $\tanh$ does not
strictly preserve the Gaussianity of the distribution of the LLR.
In this section, we consider an alternative
approach that avoids the use of the $\boxplus$ operation upon calculation of the metric associated
with check nodes by only tracking the probability where the LLR value is reversed.
We refer to this construction approach as the {\it LLR flipping probability construction}.
The resulting equations can be implemented with much less elaboration
than GA, but it still requires the assumption that 
the LLR is Gaussian distributed.
Numerical results will also reveal that such an approach
is still effective for constructing polar codes with moderate length.
We recognize that the same approach described here 
was independently proposed by Tahir and Rupp in \cite{Tahir2017}.
For the purpose of comparison, 
we briefly describe this simple and tractable approach based on our own understanding.

From the relationship of the check node given in \eqref{LN_upper},
it is easy to observe that
\begin{align}
{\rm Event}
\left[
L_{k}^{(2i)}  < 0  
\right] =
{\rm Event}
\left[
\left(
L_{k-1}^{(i)}  < 0  \cap
L_{k-1}^{(i)'}  > 0  
\right) 
\cup
\left(
L_{k-1}^{(i)}  > 0  \cap
L_{k-1}^{(i)'}  < 0  
\right) 
\right] .
\label{eq:event0}
\end{align}
Since the two events under the union operation in the right hand side of 
\eqref{eq:event0} are mutually exclusive, we may express
\begin{align}
\Pr\left(
L_{k}^{(2i)}  < 0  
\right)
&=
\Pr\left(
L_{k-1}^{(i)}  < 0  \cap
L_{k-1}^{(i)')}  > 0  
\right)
+
\Pr\left(
L_{k-1}^{(i)}  > 0  \cap
L_{k-1}^{(i)'}  < 0  
\right) \nonumber\\
& =
\Pr\left(
L_{k-1}^{(i)}  < 0  \right)
\Pr\left(
L_{k-1}^{(i)'}  > 0  
\right)
+
\Pr\left(
L_{k-1}^{(i)}  > 0  
\right) 
\Pr\left(
L_{k-1}^{(i)'}  < 0  
\right)  \nonumber \\
& =
2
\Pr\left(
L_{k-1}^{(i)}  < 0  \right)
\left[1-
\Pr\left(
L_{k-1}^{(i)}  < 0  
\right)
\right]
\label{eq:flipping}
\end{align}
where the last two equalities stem from the fact that
$L_{k-1}^{(i)}$ and $L_{k-1}^{(i)'}$ are i.i.d. 
The above equation shows that the flipping probability 
of the LLR output from the check node is
given by that of the input LLR. 
Therefore, 
the evolution of LLR through the check node
can be traced through~\eqref{eq:flipping}
if its flipping probability is of interest.
However, the LLR evaluation through the variable node, i.e.,
calculation of \eqref{LN_lower}
cannot be performed through this approach unless we know the 
distribution of the input LLR. Therefore, analogous to GA
we assume that input LLR is Gaussian distributed with mean $\mu$
and variance $\sigma^2$, i.e., $L^{(i)}_{k-1} \sim {\cal N}( \mu, \sigma^2)$.
Then, it follows that
$L^{(2i+1)}_{k} \sim {\cal N}( 2\mu, 2\sigma^2)$ and therefore
\begin{align}
 \Pr\left(
L^{(i)}_{k-1} < 0 
\right) &= Q\left(
\sqrt{\frac{ \mu^2}{\sigma^2}}
\right), \\
 \Pr\left(
L^{(2i+1)}_{k} < 0 
\right)  &= Q\left(
\sqrt{\frac{2 \mu^2}{\sigma^2}}
\right).
\end{align}
Consequently, we may write
\begin{align}
 \Pr\left(
L^{(2i+1)}_{k} < 0 
\right)  = 
Q\left[
\sqrt{2}\,
Q^{-1}\left( 
 \Pr\left(
L^{(i)}_{k-1} < 0 
\right)
\right)
\right],
\label{eq:flipping_variable}
\end{align}
where 
$Q^{-1}(\cdot): \left(0, \frac{1}{2}\right] \to [0, \infty)$ is the {\em inverse Q-function}
with $Q^{-1}\left[Q(x) \right] = x$, 
which can be calculated numerically through the bisection method.

Note that since $L^{(0)}_0 \sim {\cal N}(\gamma_0, 2 \gamma_0)$, 
we have
\begin{align}
 \Pr\left(
L^{(0)}_{0} < 0
\right)  =  
 Q\left(
\sqrt{
\frac{\gamma_0}{2}}
\right)
=
Q\left(
\sqrt{\frac{2 E_s}{N_0}}
\right),
\end{align}
which is simply the bit error rate~(BER) of uncoded BPSK as expected.

  \begin{algorithm}[t]
   \caption{Channel Polarization with LLR Flipping Probability}
   \begin{algorithmic}[1]
   \REQUIRE $n = \log_2 N$, $\alpha = {\rm SNR}_{\rm des}$ 
    \ENSURE $p[0], p[1], \ldots, p[N-1]$ as 
  $
  \Pr\left(
  L_n^{(0)} < 0
  \right),
  \Pr\left(
  L_n^{(1)} < 0
  \right),
  \cdots,
  \Pr\left(
  L_n^{(N-1)} < 0
  \right)$
    \STATE $p[0] = Q\left(
  \sqrt{2
\alpha}
  \right)$
    \FOR{$i=1$ : $\log_2 N$}
    \STATE $J = 2^{i}$
    \FOR{$j=0$ : $J/2-1$}
    \STATE $z = p[j]$
    \STATE $p[j] = 2 z ( 1 - z )$
    \STATE $p[j + J/2] = Q\left[
  \sqrt{2} Q^{-1}\left( z  \right)
  \right]$
    \ENDFOR
    \ENDFOR
    \RETURN $p[0], p[1], \cdots, p[N-1]$ 
   \end{algorithmic}
   \label{alg:proposed_flipping}
  \end{algorithm}

The channel polarization probability computation based on LLR flipping
is summarized in Algorithm~\ref{alg:proposed_flipping}.
Let $\hat{p}_i$
denote the resulting estimated
flipping probability 
${\Pr}\left(
L_n^{(i)} < 0
\right)$ 
obtained by
the algorithm with ${\rm SNR}_{\rm des}$ as its design SNR.
Then, similar to~\eqref{eq:BLER} in the case of GA, 
the minimum estimated BLER can be defined as
\begin{align}
\hat{P}_{BL,\min} 
 (R, {\rm SNR}_{\rm des}) 
&\triangleq 
\min_{{\cal I}: \,
\left|
{\cal I}\right| = R N}
\left\{
1 -
\prod_{i \in {\cal I}}
\left(
1-
\hat{p}_i
\right)
\right\} 
\label{eq:BLER3}
\end{align}
and thus selecting the 
channel indices with the smallest 
flipping probabilities
$\Pr\left(
L_n^{(i)} < 0
\right)$ may lead to the minimization of the resulting BLER.
Likewise, the estimated performance of a given designed polar code can be expressed 
in a similar form to \eqref{eq:BLER2}.

Since the function $Q(\cdot)$ returns a probability, 
it can rapidly approach $1/2$ or $0$ through the transformation
of \eqref{eq:flipping}.
To improve numerical stability, one may use a log-domain approach similar to
the improved GA described in this paper. For simplicity,
this issue will not be discussed further.
The design based on the LLR flipping probability depends on
how accurate the Gaussian modeling associated with \eqref{eq:flipping_variable} is.
It should be mentioned that the output LLR after $\boxplus$ operation
is not Gaussian in general.
Therefore, calculating the equivalent mean LLR value through the inverse
Q-function shown in \eqref{eq:flipping_variable} may not be necessarily accurate
and error may accumulate as this process is repeated.
If this part can be implemented with a more accurate model, the performance
(both BLER estimation and actual simulation based on this construction)
would be further improved.

\section{Code Design and Performance Examples}
\label{sec:example}

In this section, we investigate the performance of polar codes 
designed by the code constructions described in the previous sections
through
Monte-Carlo simulations as well as the corresponding 
BLER estimates.
Throughout all simulations, we employ the exact LLR calculations\footnote{%
Specifically, all the terms of~\eqref{eq:boxplus2} 
in {\sc Appendix}~\ref{app:meanLLR}
 are used for LLR calculation. This is in contrast to
the suboptimal min-sum decoder that calculates only the first term of~\eqref{eq:boxplus2}.}.

\subsection{Effect of Design SNR}

\begin{figure} [tbp]
\begin{center}
\includegraphics[width=.65\textwidth,clip]{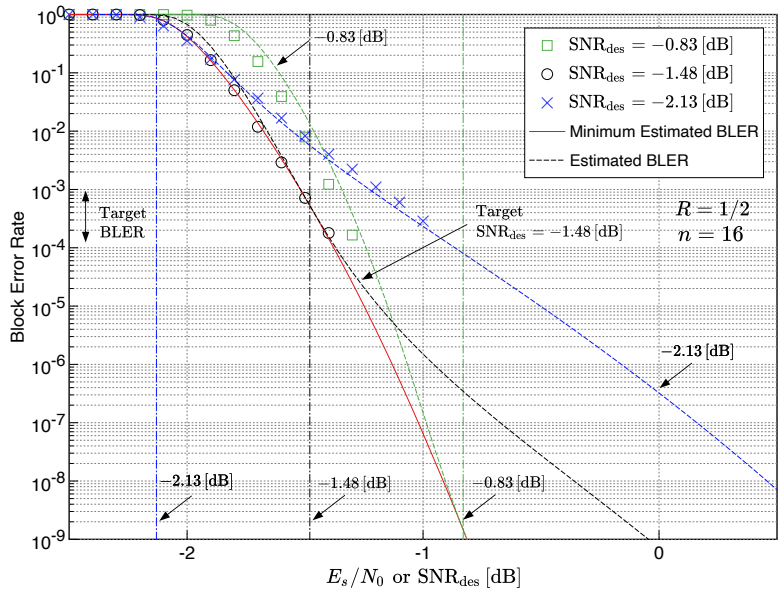}
\vspace{-.4cm}
 \caption{Comparison of minimum estimated BLER for a family of polar codes, 
estimated BLER for the polar codes
designed at specific design SNR, and the corresponding simulation results, all designed using 
Algorithm~\ref{alg:proposedGA}.
The code length is $n=16\,(N=65\,536)$ and code rate is $R=1/2$.}
\label{fig:BLER_GA_long}
\vspace{-.2cm}
\end{center}
\vspace{-.8cm}
\end{figure}

As an initial process of polar code design based on GA,
we investigate the effect of a design SNR value on the resulting 
BLER performance. To this end,
we first evaluate the minimum estimated BLER, achieved by a {\em family}
of polar codes, as a function of the design SNR based on~\eqref{eq:BLER}.
We then evaluate the estimated BLER of {\em specific} polar codes
constructed at a given design SNR, as a function of channel SNR based on~\eqref{eq:BLER2},
and compare with the corresponding simulation results.
In these steps, we employ Algorithm~\ref{alg:proposedGA},
where the inverse function corresponding to~\eqref{eq:large_gamma}
is calculated by the bisection method.
The results are shown in Fig.~\ref{fig:BLER_GA_long}
for the polar codes with code length $n=16\,(N=65\,536)$
and code rate $R=1/2$.
For the purpose of demonstration, 
we have selected three different design SNR values as follows:
We first fix the {\em target} design SNR 
as ${\rm SNR}_{\rm des} = -1.48$\,dB,
by observing where the minimum estimated BLER curve~\eqref{eq:BLER}
achieves the target BLER (set between $10^{-4}$ and $10^{-3}$).
The two other design SNR values are chosen to be either 
higher ($+0.65$\,dB) or lower ($-0.65$\,dB)
than this target value.
The comparison of the three curves clearly reveals that if the design SNR is set too low, 
the generated polar code performs better only in the low SNR region (of less practical interest), 
and {\em vice versa}. We  also notice that the polar code designed with an SNR
targeting a BLER of $10^{-3}$ could perform worse than that designed at higher SNR
targeting a lower BLER when they are compared by the required $E_s/N_0$
in order to achieve a BLER below $10^{-5}$.
This clearly demonstrates the importance of
the design SNR parameter selection
based on the target BLER, and the 
minimum estimated BLER curves serve as a good initial starting point. For the rest of the 
numerical results, we will select the design SNR such that the resulting 
minimum estimated BLER should fall between $10^{-4}$ and $10^{-3}$.

\begin{remark}
It may be worthwhile to point out the complexity required for calculating
these estimated curves.
Upon evaluating the minimum estimated BLER of~\eqref{eq:BLER},
Algorithm~\ref{alg:proposedGA} should be performed for each value of the design SNR,
which has complexity of $O(N)$.
After obtaining the estimated SNR $\hat{\gamma}^{(i)}_n$ for all $N$ channels,
the highest $K$ values should be selected for the information set ${\cal I}$
(or the lowest $N-K$ values for the frozen set ${\cal I}^c$).
The complexity of this process is $O(N\log N)$, 
which becomes non-negligible as $N$ increases.
However, the estimated SNR values of
most channels are either close to $0$ (bad channels)
or approaching $\infty$ (good channels) due to polarization.
Therefore, one can eliminate these good and bad channels 
prior to sorting, since they
are always selected as information and frozen bits, respectively.
With this pre-processing, the required time to calculate~\eqref{eq:BLER}
becomes negligible even for very long codeword cases. Once the optimal
information set ${\cal I}^*$ 
is obtained for a given design SNR,
the evaluation of the estimated BLER~\eqref{eq:BLER2} can be performed 
by running Algorithm~\ref{alg:proposedGA} again with each given channel SNR
as its input, and the resulting complexity is also negligible.
\end{remark}

\subsection{Comparison with Conventional GA and Improved GA}
\label{subsec:comparison}

\begin{figure} [tbp]
\begin{center}
\includegraphics[width=.65\textwidth,clip]{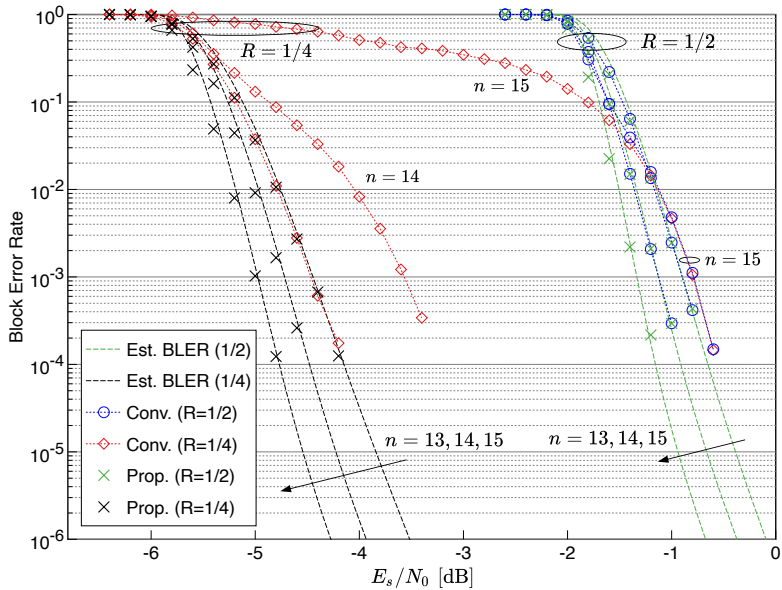}
 \caption{
Comparison of simulated BLER for the conventional and improved GA 
with code rate $R=1/2$ and $1/4$.
The code length is given by $N = 2^n$, with $n=13$, $14$, and $15$ ($N=8192$, $16\,384$, and $32\,768$). 
The corresponding estimated BLER curves as a function of $E_s/N_0$ based on the improved GA are also plotted as a reference.}
\label{fig:BLER_CONV_GA}
\end{center}
\vspace{-.8cm}
\end{figure}

We now compare the BLER performance of the 
conventional and improved GA through Monte-Carlo simulations
and demonstrate how the 
BLER of conventional GA
diverges from the estimated BLER as the code length increases.
Fig.~\ref{fig:BLER_CONV_GA} shows the simulation results of polar codes
constructed at the design SNR as described in the previous subsection.
The corresponding estimated BLER curves with respect to $E_s/N_0$ 
based on the improved GA are also plotted as a reference.
For the conventional GA, we apply \eqref{eq:chung_form}
and \eqref{eq:chung_form3},
where the bisection method is used for calculating the
inverse function of \eqref{eq:chung_form3}.
The two code rate cases of $R=1/2$ and $1/4$ are evaluated
with the code length ranging from $N=2^{13} = 8192$ to $N=2^{15}= 32\,768$.
We observe that the polar codes constructed by 
the two GA perform almost identical up to some code length as expected,
but start to diverge beyond $n=14$ for $R=1/4$, whereas
the performance of the improved GA follows similar to the estimated BLER
regardless of code length.
It is also interesting to note that for the conventional GA with $n=15$,
the two BLER curves {\em with different code rate} eventually merge, which suggests
that they share the same weak channels in the information set 
due to numerical 
inaccuracy of the conventional GA, and this inaccuracy
dominates the performance of SC decoding.

\subsection{Comparison among GA-Based Approaches and LLR Flipping Probability with $n=16$}

\begin{figure} [tbp]
\begin{center}
\includegraphics[width=.65\textwidth,clip]{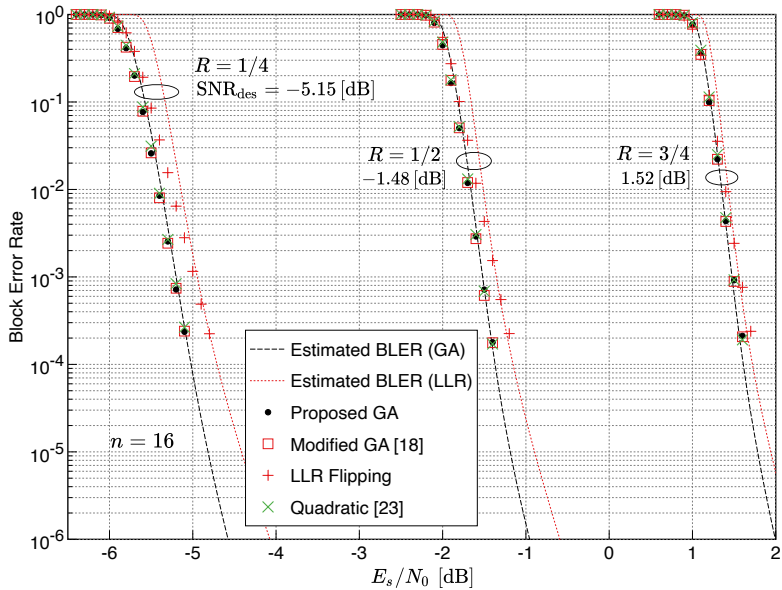}
 \caption{Comparison of BLER for the polar codes constructed by 
Algorithm~\ref{alg:proposedGA} (improved GA), 
Algorithm~\ref{alg:proposed_flipping} (LLR flipping),
the modified GA by Ha {\em et al.}~\cite{Ha2004}, and
a quadratic function approximation by Trifonov~\cite{Trifonov2018}.
The corresponding estimated BLER curves derived from
Algorithms~\ref{alg:proposedGA} and~\ref{alg:proposed_flipping} are also plotted.}
\label{fig:rate_vs_BLER512}
\end{center}
\vspace{-.8cm}
\end{figure}

Next, we compare the performances of polar codes constructed by
the GA-based approaches and the LLR flipping probability.
As an example of relatively large code length, we select $n=16$\,$(N=65\,536)$.
For construction based on the GA,
we have simulated three specific cases, i.e., the modified GA by Ha {\em et al.}~\cite{Ha2004},
the quadratic approach by Trifonov~\cite{Trifonov2018}, and the proposed GA based on 
Algorithm~\ref{alg:proposedGA}. Note that since 
the quadratic approach~\cite{Trifonov2018} does not require any calculation associated with
the inverse function of $\phi(\gamma)$ or $\xi(\gamma)$, it requires the shortest time to generate an information set 
among all the others compared here. All the polar codes compared are generated with the same
design SNR for a given code rate\footnote{%
The best design SNR that yields the minimum required $E_s/N_0$
in order to achieve a given target BLER should be different for each scheme.
However, in all the simulation results, we construct each code with 
the identical design SNR value for the purpose of demonstration.}.

The results are shown in Fig.~\ref{fig:rate_vs_BLER512} with code rates $R= 1/4$,
$1/2$, and $3/4$, where the estimated BLER
calculated by the improved GA method and that by the LLR flipping probability method are
also plotted as a reference. We observe that there is some difference in terms of the estimated BLER
between the improved GA and LLR flipping probability methods, and the gap becomes larger as the code rate decreases. The corresponding simulation results also indicate that the code 
designed by the LLR flipping probability method is worse than those constructed by 
the modification of GA method.
Therefore, even though both approaches are based on the assumption of Gaussian distribution 
for LLR, the GA-based approach may be more effective in terms of polar code design.
Furthermore, the polar codes designed by the three GA approaches exhibit similar performance.
Therefore, the difference among them is not clear from this codeword length 
even with various code rates.

\begin{figure} [tbp]
\begin{center}
\includegraphics[width=.65\textwidth,clip]{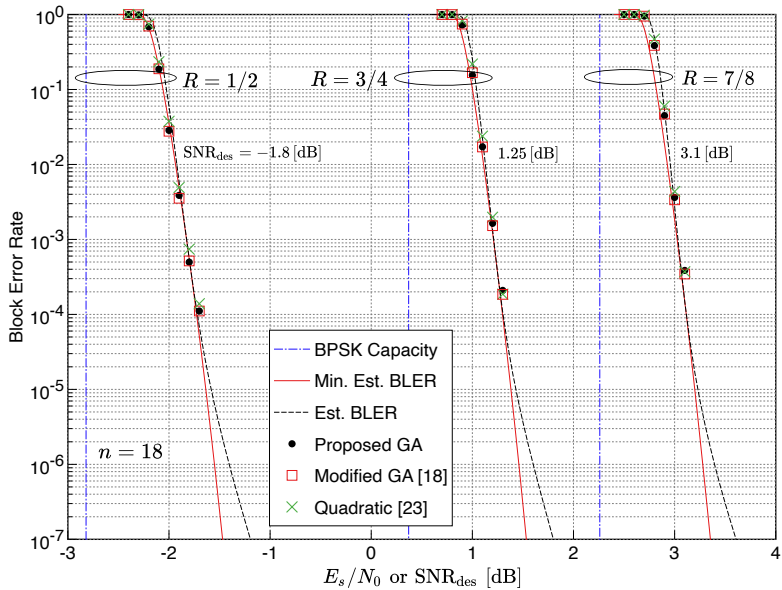}
 \caption{BLER comparison of long polar codes (with $n=18$)
based on the three GA-based construction schemes
with code rate above $1/2$. 
The minimum estimated BLER curves as a function of design SNR,
as well as the estimated BLER for each polar code simulated here, are also plotted using 
Algorithm~\ref{alg:proposedGA}.}
\label{fig:BLER_high}
\end{center}
\vspace{-.8cm}
\end{figure}

\subsection{Comparison of GA Schemes with $n=18$}

\begin{figure} [tbp]
\begin{center}
 \includegraphics[width=.65\textwidth,clip]{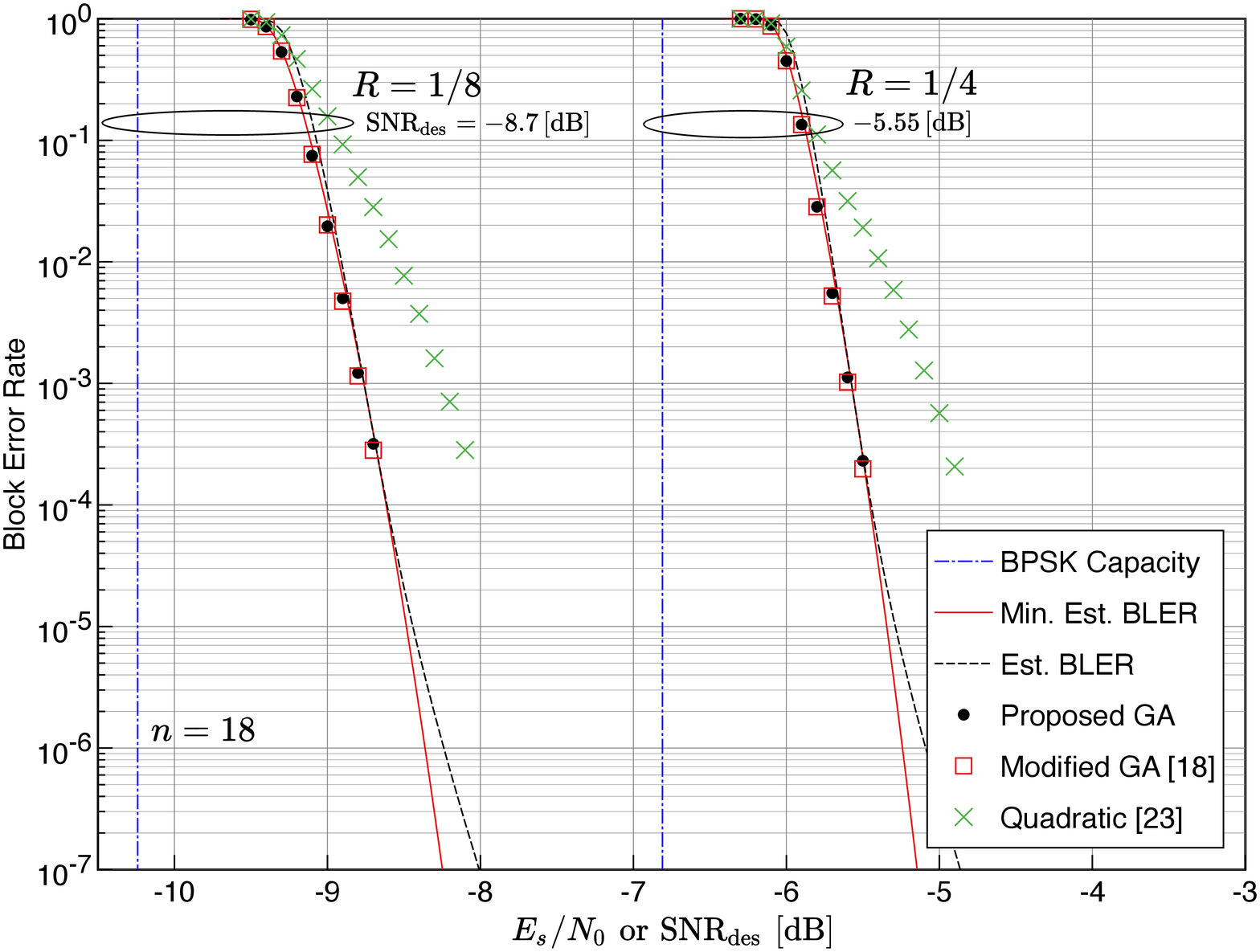}
\vspace{-.4cm}
 \caption{BLER comparison of long polar codes (with $n=18$)
based on the three GA-based construction schemes
with very low code rate. The minimum estimated BLER curves
as a function of design SNR,
as well as the estimated BLER for each polar code simulated here, are also plotted using 
Algorithm~\ref{alg:proposedGA}.}
\label{fig:BLER_low}
\vspace{-.2cm}
\end{center}
\vspace{-.8cm}
\end{figure}

Finally, we focus on even longer polar codes and evaluate the performance difference
among the three GA-based approaches. The code length is set as
$N=262\,144 \,(n=18)$ and the code rates $R$
are selected from $1/8~(0.125)$ up to $7/8~(0.875)$.
In~Fig.~\ref{fig:BLER_high}, the BLERs for the cases with higher code rates are
compared, whereas Fig.~\ref{fig:BLER_low} compares 
the BLER for the lower code rate cases.
In all the results, the minimum estimated BLER curves as a function of design SNR 
with a fixed code rate,
as well as the estimated BLER curves as a function of channel SNR, 
both employing
Algorithm~\ref{alg:proposedGA}, 
are plotted.
Also shown are the corresponding theoretical limits (the minimum $E_s/N_0$
values required to achieve the corresponding information rates in terms of the BPSK channel capacity).

From the results, we observe that the simulation results 
of the polar codes
constructed based on the proposed GA with a given design SNR
agree well with the estimated BLER evaluated at the same design SNR, especially in the BLER 
range of
practical interest ($10^{-2}-10^{-3}$).
Comparing the three GA approaches, the polar codes based on the quadratic formula
show some degradation from the other two constructions 
as the channel $E_s/N_0$ becomes lower (i.e., lower code rate cases). 
This may stem from the fact that the quadratic approximation of the function
$\Xi(\gamma)$ shows some discrepancy
from the more precise approaches as observed in Fig.~\ref{fig:mean_LLR}.
On the other hand, the proposed GA and the modified GA by Ha.~{\em et al.} 
show similar BLER performances\footnote{%
Numerical comparison has indicated that for the polar codes of $n=18$
constructed by the two GA schemes compared here, 
approximately $99.7\%$ of the frozen bits agrees when $R = 3/4$ and $7/8$,
$99.9\%$ when $R = 1/2$, 
and $99.96\%$ when $R = 1/4$ and $1/8$.}, which can also be confirmed from 
Fig.~\ref{fig:mean_LLR} as they have close resemblance even in the very 
low range of $\gamma$.

\section{Conclusions}
\label{sec:conc}

We have discussed a 
construction of
polar codes
with large code lengths based on an improved GA as well as LLR flipping probability
assuming simple SC decoding.
For the former, we have developed closed-form expressions
that allow us to precisely calculate 
the nonlinear function associated with GA 
for asymptotic cases of low and high SNR
based on its series expansion. How to select the design SNR
is a key issue for polar code design, and 
the polar codes designed using the 
estimated BLER successfully yield capacity approaching behavior as the code length increases.
Numerical results have elucidated that the GA-based design approach outperforms that
based on the LLR flipping probability.

As a final remark, even if the nonlinear transformation processes such as 
$\xi(\gamma)$ and $\xi^{-1}(\gamma)$
could be perfectly reproduced by further extending our approach, the optimality of the designed code itself is not at all guaranteed as it is based on 
an approximation. Therefore, the performance  gap from 
an optimal polar code
(that would be identified by more complex design algorithms
such as the 
approach 
developed
by Tal-Vardy) would be of significant interest.
Furthermore, the applicability of the expressions based on the log-domain analysis 
to the design of punctured and shortened polar codes would be worth investigating. 
We leave these questions for future work. 

\appendices

\section{Proof of Lemma 1}
\label{app:proof}

\begin{IEEEproof}
Equation~\eqref{eq:psi} can be expressed as
\begin{align}
 \psi(\gamma)
& =
E\left\{
\tanh\left( \frac{\sqrt{\gamma}}{2} t \right) 
\right\}, 
\end{align}
where the expectation is performed over $t \sim {\cal N}\left(\sqrt{\gamma}, 
2 \right)$.
By the series expansion of the function $\tanh(\cdot)$, we obtain~\cite{math_handbook}
\begin{align}
 \psi(\gamma)
& =
\sum_{k=1}^{\infty}
a_k 
E\left\{
\left(
\frac{\sqrt{\gamma}}{2} t
\right)^{2k - 1}
\right\}
\nonumber \\
&=
\sum_{k=1}^{\infty}
a_k
\left(
\frac{\sqrt{\gamma}}{2} 
\right)^{2k - 1}
E\left\{
 t^{2k - 1}
\right\},
\label{eq:psi_series}
\end{align}
with the coefficient $a_k$ expressed as
\begin{align}
a_k & = \frac{2^{2k}(2^k-1)}{(2k)!}B_{2k},
\label{eq:ak}
\end{align}
where $B_n$ represents the Bernolli number with its first few even number cases given by
 $B_2 =\frac{1}{6}, B_4 = -\frac{1}{30}$, and $B_6 = \frac{1}{42}$.

The odd-order moment in \eqref{eq:psi_series} can be expressed as~(see~\cite{Winkelbauer2014} for example)
\begin{align}
E\left\{
t^{2k - 1}\right\}
& = \frac{\sqrt{\gamma} }{2} \frac{(2k)!}{k!}
{}_1 F_1\left( 1- k; \frac{3}{2}; - \frac{\gamma}{4}
\right),
\label{eq:moment}
\end{align}
where ${}_1 F_1(a;b;x)$ is Kummer's confluent hypergeometric function.
Substituting
\eqref{eq:ak} and \eqref{eq:moment} into
\eqref{eq:psi_series} and after some manipulation, we obtain
\begin{align}
 \psi(\gamma)
& = 
\sum_{k=1}^{\infty}
\frac{
\left(
2^{2k}-1
\right)
 B_{2k}
}{k!}
\gamma^k 
{}_1 F_1\left( 1- k; \frac{3}{2}; - \frac{\gamma}{4}
\right).
\label{eq:series_final}
\end{align}
Note that the function ${}_1 F_1(a;b;x)$ is differentiable with respect to $x$
and converges for any finite $x$.
Since ${}_1F_1(0;b;x) = 1$, ${}_1F_1(-1;b;x) = 1-\frac{x}{b}$, and 
${}_1F_1(-2;b;x) = 1-\frac{2x}{b} + \frac{x^2}{b(1+b)}$, 
by expanding the terms of~\eqref{eq:series_final} up to $k=3$, we
obtain \eqref{eq:th01}.
\end{IEEEproof}

\section{Mean of LLR Output}
\label{app:meanLLR}

In this appendix, we derive the mean value of the LLR output 
associated with the operation $L_o = L_a \boxplus L_b$,
assuming that the two input LLRs $L_a$ and $L_b$
are independent Gaussian with ${\cal N}(\gamma, 2 \gamma)$.

We first note that the output $L_o$ associated
with the  $\boxplus$ operation of two LLRs $L_a$ and $L_b$
can be expressed 
by the so-called min-sum form:
\begin{align}
L_o & \triangleq 
L_a \boxplus L_b \nonumber\\
& =
\underbrace{%
\text{sign}
\left(
L_a
\right)
\text{sign}
\left(
L_b
\right)
\min
\left(
\left|
L_a
\right|,
\left|
L_b
\right|
\right)}_{\triangleq X}
+ 
\log \left(
1+e^{-\left|
L_a + L_b
\right|}
\right)
-
\log \left(
1+e^{-\left|
L_a - L_b
\right|}
\right),
\label{eq:boxplus2} 
\end{align}
where $X$ corresponds to the output of the min-sum decoder, and is often
used as a low-complexity alternative to exact decoding.
As discussed in~\cite{Kern2014},
based on the fundamental results on order statistics,
the probability density function~(pdf) of the random variable $X$ defined above can be expressed as
\begin{align}
 f_X(x) & = 2 f_{L}(x) \left\{
1 - F_L\left(\left|x\right|\right)
\right\} + 2 f_L(-x)
 F_L\left(- \left|x\right|\right),
\end{align}
where $f_L(x)$ and $F_L(x)$ are the pdf and 
cumulative distribution function~(cdf) of $L_a$ (or equivalently $L_b$), respectively.
Assuming $L_a, L_b \sim {\cal N}(\gamma, 2 \gamma)$,
the above pdf can be expressed as
\begin{align}
f_X(x) = \frac{1}{\sqrt{\pi \gamma}}
\left\{
e^{- \frac{
\left(
x - \gamma
\right)^2
}
{4\gamma}} 
Q\left(
\frac{\left|x\right| - \gamma 
}{\sqrt{2 \gamma}}
\right)
+ 
e^{- \frac{
\left(
x + \gamma
\right)^2
}
{4\gamma}} 
 Q\left(
\frac{\left|x\right| + \gamma 
}{\sqrt{2 \gamma}}
\right)
\right\} .
\label{eq:fX}
\end{align}

Next, let us define the random variables 
$Y = \left|L_a + L_b\right|$
and $Z = \left|L_a - L_b\right|$ which appear in~\eqref{eq:boxplus2}.
Since $L_a + L_b \sim {\cal N}(2\gamma, 4\gamma)$
and $L_a - L_b \sim {\cal N}(0, 4\gamma)$,
$Y$ follows a {\em generalized} Rice distribution and $Z$
follows  a {\em generalized} Rayleigh distribution~\cite{proakis08},
whose pdfs are expressed as
\begin{align}
 f_Y(y) & = 
\frac{1}{\sqrt{{2 \pi \gamma}}} 
e^{-\frac{y^2 + 4 \gamma^2}{8 \gamma}}
\cosh\left(
\frac{y}{2}
\right) \\
f_Z(z) & = \frac{1}{\sqrt{{2 \pi \gamma}}} 
e^{-\frac{z^2}{8 \gamma}}.
\end{align}
Consequently,
the mean of $L_o$ can be expressed as
\begin{align}
 E\left[
L_o
\right] & = 
E\left[X\right]
+
E\left[
\log \left(
1+e^{-Y}\right)
\right]
-
 E\left[
\log \left(
1+e^{-Z}
\right)
\right] \nonumber \\
& = 
\frac{1}{\sqrt{\pi \gamma}}
\int_{0}^{\infty}
\left[
z e^{-\frac{(z - \gamma)^2}{4\gamma}} 
\left(
1 - e^{-z}
\right)
\left[
 Q\left(
\frac{z - \gamma}{\sqrt{2 \gamma}}
\right)
-
 Q\left(
\frac{z + \gamma}{\sqrt{2 \gamma}}
\right)
\right]\right. \nonumber \\
& \left. \qquad - \frac{1}{\sqrt{2}} \log\left(
1 + e^{-z} 
\right)\left\{
e^{-\frac{z^2}{8\gamma}}
- \frac{1}{2}
e^{-\frac{1}{8\gamma} \left(z - 2\gamma\right)^2}
\left(1 + e^{-z}\right)
\right\}
\right] d z 
\label{eq:appendix_meanLLR0}
\\
& = 
\frac{1}{\sqrt{\pi}}
\int_{0}^{\infty}
\left[
\sqrt{\gamma}
x
e^{-\frac{(x - \sqrt{\gamma} )^2}{4}} 
\left(
1 - e^{-\sqrt{\gamma} x}
\right)
\left[
 Q\left(
\frac{x - \sqrt{\gamma} }{\sqrt{2}}
\right)
-
 Q\left(
\frac{x + \sqrt{\gamma} }{\sqrt{2}}
\right)
\right]\right. \nonumber \\
& \left. \qquad - \frac{1}{\sqrt{2}} \log\left(
1 + e^{-\sqrt{\gamma} x} 
\right)\left\{
e^{-\frac{x^2}{8}}
- \frac{1}{2}
e^{-\frac{1}{8} \left( 
x
- 2\sqrt{\gamma} \right)^2}
\left(1 + e^{-
\sqrt{\gamma} x
}\right)
\right\}
\right] 
dx ,
\label{eq:appendix_meanLLR}
\end{align}
where \eqref{eq:appendix_meanLLR} may be convenient for small values of
$\gamma$ and \eqref{eq:appendix_meanLLR0} is suitable as $\gamma$ increases
for numerical evaluation.

\section*{Acknowledgment}
\addcontentsline{toc}{section}{Acknowledgment}
The authors would like to thank the editor and reviewers for their constructive comments 
as well as bringing~\cite{Ha2004,Trifonov2018} to the authors' attention.

\end{document}